\documentclass[11pt]{article}
\usepackage{epsfig} 
\usepackage{amssymb}
\usepackage{float}
\newcommand{\sect}[1]{ \section{#1} \setcounter{equation}{0} }

\newcommand{\half}{\mbox{\small{$\frac{1}{2}$}}}

\newcommand{\MSbar}{\overline{\mbox{MS}}}

\begin{document}
\title{Six loop critical exponent analysis for Lee-Yang and percolation theory}

\author{J.A. Gracey, \\ Theoretical Physics Division, \\ 
Department of Mathematical Sciences, \\ University of Liverpool, \\ P.O. Box 
147, \\ Liverpool, \\ L69 3BX, \\ United Kingdom.} 
\date{} 

\maketitle 

\vspace{5cm} 
\noindent 
{\bf Abstract.} Using the recent six loop renormalization group functions for
Lee-Yang and percolation theory constructed by Schnetz from a scalar cubic
Lagrangian, we deduce the $\epsilon$ expansion of the critical exponents for
both cases. Estimates for the exponents in three, four and five dimensions are 
extracted using two-sided Pad\'{e} approximants and shown to be compatible with 
values from other approaches.

\vspace{-16.0cm}
\hspace{13.4cm}
{\bf LTH 1409}

\newpage

\sect{Introduction.}

In recent years there has been a significant advance in the renormalization of
core quantum field theories to high loop order due to the development of new
techniques such as the Laporta algorithm, \cite{1}, the {\sc Forcer} package,
\cite{2,3}, high numerical precision methods based on difference equations to
determine massive tadpole graphs, \cite{1}, and graphical functions \cite{4,5}. 
These tools have opened the door to $\beta$-functions to five loops and beyond 
\cite{6,7,8,9,10,11,12,13,14}. While \cite{6,7,8} have extended the
renormalization of Quantum Chromodynamics (QCD) to five loops and that of 
Gross-Neveu-Yukawa theory to the same order, \cite{9}, renormalization at
orders beyond five have also been achieved with the six loop $\beta$-function 
of scalar $\phi^3$ theory available now in \cite{14} and the seven loop 
$\phi^4$ theory $\beta$-function provided in \cite{11}. Such precision for 
gauge theories, for instance, has refined uncertainties in observables for 
particle physics experiments as well as in applications to condensed matter 
physics based on scalar and Gross-Neveu-Yukawa theories. More specifically for 
the latter critical exponent estimates governing the properties of phase 
transitions have been refined. 

Of the suite of theories mentioned earlier the six loop renormalization group
functions of scalar $\phi^3$ theory have been established most recently in 
\cite{14} using graphical functions. As a by-product the renormalization group 
functions for Lee-Yang and percolation theory were deduced to the same order in
the modified minimal subtraction ($\MSbar$) scheme, \cite{15}. For instance the
connection of the Lee-Yang singularity problem with scalar $\phi^3$ theory was
originally made in \cite{16}. The physics these models relate to are important 
in condensed matter with critical exponents calculated via a variety of methods 
such as high temperature and Monte Carlo techniques in fixed dimensions between
two and six where the latter is the critical dimension of a scalar theory with 
a cubic interaction. More modern approaches such as the conformal bootstrap 
programme and functional renormalization group techniques have also been used 
to study both problems. We note that Lee-Yang theory also has applications to 
particle theory due to the relation of the associated edge singularity problem 
with the QCD equation of state \cite{17}. This relies on an accurate value of 
the exponent $\sigma$ in three dimensions. Another example where $\sigma$ is 
important arose in a recent QCD lattice analysis relating to baryon number, 
\cite{18}. There the volume scaling of Lee-Yang zeroes, when there is crossover
behaviour, is determined by $\sigma$. For a deeper overview of these 
connections see \cite{19}.

The other technique used to determine exponent estimates is that of the
$\epsilon$ expansion. It is based on the cubic scalar theory renormalization 
group functions in $d$~$=$~$6$~$-$~$\epsilon$ spacetime dimensions with the
critical exponents determined as a power series in $\epsilon$ at the 
Wilson-Fisher fixed point, \cite{20}. For the Lee-Yang and percolation
situations the exponents were produced over many years from low order up to
five loops \cite{12,13,21,22,23,24}. Compared with results known at each 
respective time from other techniques exponent estimates derived from the
$\epsilon$ expansion were in solid agreement for three, four and five
dimensions. This may appear odd since to reach say three dimensions from six 
one would have to set $\epsilon$~$=$~$3$ thereby raising the question of
convergence. In \cite{13,24} exponents were deduced numerically by several 
approaches one of which was that of two-sided Pad\'{e} approximants. While one 
side was the exponent value in strictly six dimensions there were lower 
dimensional boundary conditions for each exponent. These were the respective 
values of the exponents in two dimensions derived from the separate conformal 
field theories that govern transitions in the Lee-Yang or percolation cases.
The values of critical exponents of such field theories were determined 
{\em exactly} using conformal symmetry in \cite{25}. For the Lee-Yang model 
exponent values are also known exactly in one dimension. So in computing 
two-sided approximants in \cite{13,24} those for Lee-Yang used two constraints 
while only one was available for percolation theory. 

Given this background it is therefore the purpose of this article to extend the
results of \cite{13,22,23,24} to six loops given the advance made in 
\cite{14,15}. With the extra order and the lower dimension constraints more 
Pad\'{e} approximants are available which has allowed us to provide an 
uncertainty measure on the estimates we present here. This is qualified of 
course by the fact that a rational polynomial approximation to a continuous 
function of $d$ in $2$~$\leq$~$d$~$\leq$~$6$ may not itself be continuous. This
transpired to be the case in that several of the approximants which were 
constructed  had singularities or were not monotonic. The last criterion is 
derived from the behaviour of available estimates from other approaches meaning
that in $2$~$\leq$~$d$~$\leq$~$6$ they were either increasing or decreasing
monotonically. Subject to these caveats we managed to obtain exponent estimates
for both Lee-Yang and percolation theory that were credible compared with other
techniques. Moreover they were not significantly different from previous lower 
loop values thereby indicating a degree of convergence as the loop order 
increases.

The article is structured as follows. The focus in Section $2$ is on Lee-Yang
theory where we record the $\epsilon$ expansion of the key critical exponents
before detailing the two-sided Pad\'{e} formalism. Exponent estimates using
this method and their comparison with known values complete the section. The
parallel analysis for percolation theory is given in Section $3$ followed by
concluding remarks in Section $4$.

\sect{Lee-Yang exponents.}

For the first part of the analysis we concentrate on the Lee-Yang problem which
is underpinned by a non-unitary Lagrangian with a single scalar field $\phi$
self-interacting cubically with coupling $g$ obeying the Lagrangian
\begin{equation}
L ~=~ \frac{1}{2} \left( \partial_\mu \phi \right)^2 ~+~ \frac{ig}{6} \phi^3
\label{lagphi3}
\end{equation}
with the relation to the Lee-Yang edge singularity being elucidated in
\cite{16}. En route we will illustrate the approach to our analysis which will 
also be applied to percolation theory. As the six loop $\MSbar$ 
$\beta$-function for (\ref{lagphi3}) as well as the $\phi$ field anomalous 
dimension were recorded in \cite{14,15} their associated critical exponents are 
straightforward to extract at the Wilson-Fisher critical point, $g^\star$, in 
$d$~$=$~$6$~$-$~$\epsilon$ as
\begin{eqnarray}
\hat{\eta} &=&
-~ \frac{1}{9} \epsilon
- \frac{43}{729} \epsilon^2
+ \left[
\frac{16}{243} \zeta_3
- \frac{8375}{236196}
\right] \epsilon^3
+ \left[
\frac{4}{81} \zeta_4
- \frac{3883409}{76527504}
- \frac{716}{19683} \zeta_3
- \frac{80}{2187} \zeta_5
\right] \epsilon^4
\nonumber \\
&&
+ \left[
\frac{136}{2187} \zeta_3^2
+ \frac{37643}{59049} \zeta_3
+ \frac{80524}{59049} \zeta_5
- \frac{1545362585}{24794911296}
- \frac{4172}{2187} \zeta_7
- \frac{179}{6561} \zeta_4
- \frac{100}{2187} \zeta_6
\right] \epsilon^5
\nonumber \\
&&
+ \left[
\frac{68}{729} \zeta_3 \zeta_4
+ \frac{3584}{6561} \zeta_3^3
+ \frac{4624}{3645} \zeta_{5,3}
+ \frac{37643}{78732} \zeta_4
+ \frac{99440}{59049} \zeta_6
+ \frac{129968}{177147} \zeta_3^2
+ \frac{1734712}{59049} \zeta_9
\right. \nonumber \\
&& \left. ~~~~
-~ \frac{152041988501}{2677850419968}
- \frac{440890397}{258280326} \zeta_3
- \frac{158942329}{28697814} \zeta_5
- \frac{6992407}{354294} \zeta_7
- \frac{15616}{59049} \zeta_3 \zeta_5
\right. \nonumber \\
&& \left. ~~~~
-~ \frac{7781}{1215} \zeta_8
\right] \epsilon^6 ~+~ O(\epsilon^7)
\nonumber \\
\hat{\omega} &=&
\epsilon
- \frac{125}{162} \epsilon^2
+ \left[
\frac{20}{27} \zeta_3
+ \frac{36755}{52488}
\right] \epsilon^3
+ \left[
\frac{5}{9} \zeta_4
+ \frac{160}{81} \zeta_5
- \frac{31725355}{17006112}
- \frac{9673}{2187} \zeta_3
\right] \epsilon^4
\nonumber \\
&&
+ \left[
\frac{17088604709}{5509980288}
- \frac{21980}{243} \zeta_7
- \frac{9673}{2916} \zeta_4
+ \frac{200}{81} \zeta_6
+ \frac{1384}{243} \zeta_3^2
+ \frac{1050770}{19683} \zeta_5
+ \frac{12094613}{354294} \zeta_3
\right] \epsilon^5
\nonumber \\
&&
+ \left[
\frac{692}{81} \zeta_3 \zeta_4
+ \frac{5968}{81} \zeta_{5,3}
+ \frac{23680}{729} \zeta_3^3
+ \frac{1880255}{59049} \zeta_3^2
+ \frac{5235625}{78732} \zeta_6
+ \frac{9801800}{6561} \zeta_9
+ \frac{12094613}{472392} \zeta_4
\right. \nonumber \\
&& \left. ~~~~
-~ \frac{3218943170927}{595077871104}
- \frac{22622670455}{114791256} \zeta_3
- \frac{677377966}{1594323} \zeta_5
- \frac{14402270}{19683} \zeta_7
- \frac{81733}{243} \zeta_8
\right. \nonumber \\
&& \left. ~~~~
-~ \frac{16640}{2187} \zeta_3 \zeta_5
\right] \epsilon^6 ~+~ O(\epsilon^7) 
\end{eqnarray}
where $\hat{\eta}$~$=$~$\gamma_\phi(g^\star)$,
$\hat{\omega}$~$=$~$\beta^\prime(g^\star)$, $\zeta_n$ is the Riemann zeta 
function and $\zeta_{5,3}$~$=$~$\sum_{m>n\geq 1} \frac{1}{m^5 n^3}$ is a 
multiple zeta. For this section our notation is to place a hat on the exponents
connected with Lee-Yang theory to distinguish them from similar exponents we 
will compute in percolation theory. In addition to $\hat{\eta}$ and 
$\hat{\omega}$ there are several other exponents that are important for 
physical problems and which can be derived from scaling laws \cite{26}. These 
are, \cite{26},
\begin{eqnarray}
\hat{\sigma} &=& \frac{[d-2+\hat{\eta}]}{[d+2-\hat{\eta}]} ~~~,~~~
\hat{\phi} ~=~ 1 ~+~ \hat{\sigma} ~~~,~~~
\hat{\theta} ~=~ \hat{\nu}_c \hat{\omega} \nonumber \\
\hat{\nu} &=& \frac{2}{[d-2+\hat{\eta}]} ~~~,~~~
\hat{\nu}_c ~=~ \frac{2}{[d+2-\hat{\eta}]}
\label{scalsigma}
\end{eqnarray}
where $\hat{\theta}$ and $\hat{\omega}$ are correction to scaling exponents and
$\hat{\nu}$ and $\hat{\nu}_c$ are not independent since \cite{27}
\begin{equation}
\frac{1}{\hat{\nu}} ~+~ \frac{1}{\hat{\nu}_c} ~=~ d ~.
\end{equation}
In addition it is known that the exponent $\hat{\beta}$ is unity in all
dimensions $d$ \cite{22,27}. Consequently the $\epsilon$ expansion of the 
exponents we will find estimates for are
\begin{eqnarray}
\hat{\sigma} &=&
\frac{1}{2}
- \frac{1}{12} \epsilon
- \frac{79}{3888} \epsilon^2
+ \left[
\frac{1}{81} \zeta_3
- \frac{10445}{1259712}
\right] \epsilon^3
+ \left[
\frac{1}{108} \zeta_4
- \frac{4047533}{408146688}
- \frac{161}{26244} \zeta_3
- \frac{5}{729} \zeta_5
\right] \epsilon^4
\nonumber \\
&&
+ \left[
\frac{17}{1458} \zeta_3^2
+ \frac{20101}{78732} \zeta_5
+ \frac{112399}{944784} \zeta_3
- \frac{1601178731}{132239526912}
- \frac{1043}{2916} \zeta_7
- \frac{161}{34992} \zeta_4
- \frac{25}{2916} \zeta_6
\right] \epsilon^5
\nonumber \\
&&
+ \left[
\frac{17}{972} \zeta_3 \zeta_4
+ \frac{224}{2187} \zeta_3^3
+ \frac{289}{1215} \zeta_{5,3}
+ \frac{32669}{236196} \zeta_3^2
+ \frac{49645}{157464} \zeta_6
+ \frac{112399}{1259712} \zeta_4
+ \frac{216839}{39366} \zeta_9
\right. \nonumber \\
&& \left. ~~~~
-~ \frac{158574097133}{14281868906496}
- \frac{156752701}{153055008} \zeta_5
- \frac{107952203}{344373768} \zeta_3
- \frac{7029955}{1889568} \zeta_7
- \frac{7781}{6480} \zeta_8
\right. \nonumber \\
&& \left. ~~~~
-~ \frac{976}{19683} \zeta_3 \zeta_5
\right] \epsilon^6 ~+~ O(\epsilon^7) \nonumber \\
\hat{\nu}_c &=&
\frac{1}{4}
+ \frac{1}{36} \epsilon
+ \frac{29}{23328} \epsilon^2
+ \left[
\frac{1}{486} \zeta_3
- \frac{8879}{7558272}
\right] \epsilon^3
\nonumber \\
&&
+ \left[
\frac{1}{648} \zeta_4
- \frac{4526999}{2448880128}
- \frac{107}{157464} \zeta_3
- \frac{5}{4374} \zeta_5
\right] \epsilon^4
\nonumber \\
&&
+ \left[
\frac{17}{8748} \zeta_3^2
+ \frac{20011}{472392} \zeta_5
+ \frac{111757}{5668704} \zeta_3
- \frac{1845636677}{793437161472}
- \frac{1043}{17496} \zeta_7
- \frac{107}{209952} \zeta_4
\right. \nonumber \\
&& \left. ~~~~
-~ \frac{25}{17496} \zeta_6
\right] \epsilon^5
\nonumber \\
&&
+ \left[
\frac{17}{5832} \zeta_3 \zeta_4
+ \frac{112}{6561} \zeta_3^3
+ \frac{289}{7290} \zeta_{5,3}
+ \frac{4141}{177147} \zeta_3^2
+ \frac{12355}{236196} \zeta_6
+ \frac{111757}{7558272} \zeta_4
+ \frac{216839}{236196} \zeta_9
\right. \nonumber \\
&& \left. ~~~~
-~ \frac{191795557319}{85691213438976}
- \frac{404651861}{8264970432} \zeta_3
- \frac{150269137}{918330048} \zeta_5
- \frac{7142599}{11337408} \zeta_7
- \frac{7781}{38880} \zeta_8
\right. \nonumber \\
&& \left. ~~~~
-~ \frac{488}{59049} \zeta_3 \zeta_5
\right] \epsilon^6 ~+~ O(\epsilon^7) \nonumber \\
\hat{\theta} &=&
\frac{1}{4} \epsilon
- \frac{107}{648} \epsilon^2
+ \left[
\frac{5}{27} \zeta_3
+ \frac{8129}{52488}
\right] \epsilon^3
+ \left[
\frac{5}{36} \zeta_4
+ \frac{40}{81} \zeta_5
- \frac{3818417}{8503056}
- \frac{9475}{8748} \zeta_3
\right] \epsilon^4
\nonumber \\
&&
+ \left[
\frac{7972456399}{11019960576}
- \frac{9475}{11664} \zeta_4
- \frac{5495}{243} \zeta_7
+ \frac{50}{81} \zeta_6
+ \frac{346}{243} \zeta_3^2
+ \frac{263750}{19683} \zeta_5
+ \frac{11918591}{1417176} \zeta_3
\right] \epsilon^5
\nonumber \\
&&
+ \left[
\frac{173}{81} \zeta_3 \zeta_4
+ \frac{1492}{81} \zeta_{5,3}
+ \frac{5920}{729} \zeta_3^3
+ \frac{959221}{118098} \zeta_3^2
+ \frac{2450450}{6561} \zeta_9
+ \frac{5256775}{314928} \zeta_6
+ \frac{11918591}{1889568} \zeta_4
\right. \nonumber \\
&& \left. ~~~~
-~ \frac{377937484105}{297538935552}
- \frac{11090116969}{229582512} \zeta_3
- \frac{1335259205}{12754584} \zeta_5
- \frac{29209567}{157464} \zeta_7
- \frac{81733}{972} \zeta_8
\right. \nonumber \\
&& \left. ~~~~
-~ \frac{4160}{2187} \zeta_3 \zeta_5
\right] \epsilon^6 ~+~ O(\epsilon^7) ~.
\end{eqnarray}
To gauge how each series behaves we note that the analytic expressions
translate into
\begin{eqnarray}
\hat{\eta} &=&
-~ 0.111111 \epsilon    
- 0.058985 \epsilon^2     
+ 0.043690 \epsilon^3
- 0.078954 \epsilon^4     
+ 0.208254 \epsilon^5     
- 0.667225 \epsilon^6 ~+~ O(\epsilon^7) \nonumber \\
\hat{\sigma} &=& 0.500000
- 0.083333 \epsilon           
- 0.020319 \epsilon^2           
+ 0.006549 \epsilon^3            
- 0.014382 \epsilon^4          
+ 0.038113 \epsilon^5           
- 0.122697 \epsilon^6 
\nonumber \\
&& +~ O(\epsilon^7) \nonumber \\
\hat{\nu}_c &=&
0.250000 
+ 0.027778 \epsilon 
+ 0.001243 \epsilon^2 
+ 0.001299 \epsilon^3 
- 0.002180 \epsilon^4 
+ 0.005989 \epsilon^5 
- 0.019451 \epsilon^6
\nonumber \\
&& +~ O(\epsilon^7) \nonumber \\
\hat{\theta} &=& 0.250000 \epsilon
- 0.165123 \epsilon^2
+ 0.377477 \epsilon^3
- 1.088632 \epsilon^4
+ 3.731811 \epsilon^5
- 14.380734 \epsilon^6 ~+~ O(\epsilon^7) \nonumber \\
\hat{\omega} &=& \epsilon
- 0.771605 \epsilon^2
+ 1.590668 \epsilon^3
- 4.532626 \epsilon^4
+ 15.435688 \epsilon^5
- 59.254423 \epsilon^6 ~+~ O(\epsilon^7)
\label{lyexp}
\end{eqnarray}
numerically. We note that a numerical prediction was provided for $\hat{\phi}$ 
to $O(\epsilon^6)$ in \cite{26}. While its terms to $O(\epsilon^3)$ used the 
then known three loops results of \cite{22,23} the subsequent terms to 
$O(\epsilon^6)$ were found by constructing a constrained $[6,0]$ Pad\'{e}
approximant from estimates of $\hat{\phi}$ in $d$~$=$~$3$, $4$ and $5$
dimensions. The coefficients of the resulting higher order term in $\epsilon$
are significantly smaller than the corresponding ones of $\hat{\sigma}$ in 
(\ref{lyexp}) when the scaling relation between $\hat{\phi}$ and $\hat{\sigma}$
is employed.

In previous perturbative analyses of critical exponents in $\phi^3$ theory one
method of extracting estimates was to use Pad\'{e} approximants and we follow
the same approach here. While the Pad\'{e} method uses rational polynomials to 
approximate a function in a parameter which is regarded as small there is no
guarantee that an approximant will capture the salient features of the exact
function. For instance if the function is continuous an approximant may 
contain discontinuities or if the function is monotonic the Pad\'{e} may have
stationary points. In the present situation the aim is to create approximants
that are valid in $2$~$\leq$~$d$~$\leq$~$6$ and then read off estimates for
three, four and five dimensions. The obvious concern is that while $\epsilon$
is regarded as small in relation to the establishment of the Wilson-Fisher
fixed point perturbatively it would be unlikely to be applicable when 
$\epsilon$~$=$~$3$. However in previous lower loop analyses the construction
of Pad\'{e} approximants for Lee-Yang and percolation theory exponents 
benefitted greatly from knowledge of the {\em exact} exponents in two
dimensions in both instances. These have been established from connecting the
critical theory in $d$~$\neq$~$2$ to a known conformal field theory in two
dimensions which were all classified in \cite{25}. In the Lee-Yang case exact 
exponents are also available in one dimension. Therefore we will use such
available information to construct what is termed two-sided or constrained 
Pad\'{e} approximants for each of the above exponents as well as those for
percolation. 

{\begin{table}[ht]
\begin{center}
\begin{tabular}{|c||c|c|c|c|c|c|}
\hline
\rule{0pt}{15pt}
$d$ & $\hat{\eta}$ & $\hat{\sigma}$ & $\hat{\nu}_c$ & $\hat{\theta}$ & 
$\hat{\omega}$ & $\Delta_{\phi \frac{}{}}$ \\
\hline
\rule{0pt}{15pt}
$1$ & $-~ 1$ & $-~ \frac{1}{2}$ & $\frac{1}{2}$ & $(1)$ & $(2)$ & $-~ 1$ \\
\rule{0pt}{15pt}
$2$ & $-~ \frac{4}{5}$ & $-~ \frac{1}{6}$ & $\frac{5}{12}$ & $\frac{5}{6}$ & $2$ & $-~ \frac{2}{5}$ \\
\rule{0pt}{15pt}
$6$ & $0$ & $\frac{1}{2}_{\frac{}{}}$ & $\frac{1}{4}_{\frac{}{}}$ & $0$ & $0$ & $2$ \\
\hline
\end{tabular}
\caption{Values of Lee-Yang exponents in one, two and six dimensions.}
\label{lyd126}
\end{center}
\end{table}}

One advantage of using the lower dimensional constraints is that the 
approximants should approach a more reliable value in three dimensions. Looking
at it from another point of view not using the constraints for a Pad\'{e} 
exponent could give a value well away from the exact value in two dimensions
with an indication of this provided in \cite{24}. We have recorded the values 
of the exponents in low dimensions in Table \ref{lyd126} where knowing 
$\hat{\eta}$ in one and two dimensions determines $\hat{\sigma}$ and 
$\hat{\nu}_c$ whereas $\hat{\theta}$ for two dimensions was taken from 
\cite{26} and it determines $\hat{\omega}$. In $d$~$=$~$1$ the values for 
$\hat{\theta}$ and $\hat{\omega}$ are from \cite{26,28,29} but are bracketed 
since a doubt was expressed in \cite{26} as to whether the values were reliable
due to possible logarithmic behaviour in these correction to scaling exponents.
Though it was noted in \cite{26} that the values we have bracketed may be valid
in the limit $d$~$\to$~$1^+$. Given this we have used a cautious approach in 
constructing Pad\'{e} approximants for $\hat{\theta}$ and $\hat{\omega}$ by 
using the two dimensional constraint only. One comment in relation to the 
distinction between approximants with one and two constraints is worth noting. 
This is that the use of a $d$~$=$~$1$ value informs the {\em gradient} of the 
approximant in the region above two dimensions rather than solely the actual 
boundary point. In influencing the slope it refines the direction of the 
approximant towards three dimensions. The final exponent in Table \ref{lyd126},
$\Delta_\phi$, relates to the full dimension of the $\phi$ field which is 
primarily used in the conformal bootstrap formalism. We include it in our 
discussions in order to translate exponent estimates from that and other 
techniques to the remaining exponents in the table. For completeness we note 
the relations are, \cite{30},
\begin{equation}
\hat{\eta} ~=~ 2 ~-~ d ~+~ 2 \Delta_\phi ~~~,~~~ 
\hat{\sigma} ~=~ \frac{\Delta_\phi}{[d-\Delta_\phi]} ~~~,~~~
\hat{\nu}_c ~=~ \frac{1}{[d-\Delta_\phi]} 
\label{scaldelta}
\end{equation}
and
\begin{equation}
\hat{\omega} ~=~ -~ d ~+~ \Delta_{\phi^3} 
\end{equation}
where $\Delta_{\phi^3}$ is the full dimension of the spin-$0$ conformal primary
field $\phi^3$ which is present in (\ref{lagphi3}).

In using the two-sided Pad\'{e} approach for lower loops, \cite{13,24}, the
results for exponents in three dimensions were generally in line with estimates
from other techniques. The aim is that with the inclusion of six loop 
information uncertainties can be refined. This should be possible if one 
considers the data available for an $L$ loop exponent. Including the canonical 
dimension means there are $(L+1)$ coefficients in the $\epsilon$ expansion of
an exponent. Including two constraints from lower dimensions gives a total of
$(L+3)$ pieces of information to construct the constrained Pad\'{e} 
approximants. Allowing for a normalization drops this to $(L+2)$ data points to
construct the set of rational polynomials at $L$ loops. Excluding the $[p,0]$ 
Pad\'{e}s means at $L$ loops there are $(L+2)$ possible approximants where the 
$[p/q]$ approximant, ${\cal P}_{[p/q]}(\epsilon)$, is defined by
\begin{equation}
{\cal P}_{[p/q]}(\epsilon) ~=~
\frac{\sum_{m=0}^p a_m \epsilon^m}{1 + \sum_{n=1}^q b_n \epsilon^n} ~.
\end{equation}
So at six loops there will be eight Pad\'{e}s to include in any analysis. For 
exponents where there is only one constraint from low dimension there will be 
$(L+1)$ possible approximants. If we construct approximants starting at two 
loops this means the total available number of approximants will be 
$\half (L-1) (L+4)$ and $\half (L-1) (L+6)$ when one and two constraints are 
employed respectively. When the leading term of the $\epsilon$ expansion is 
$O(\epsilon)$ these numbers reduce by unity at each loop order. Aside from this
caveat this ought to improve uncertainty estimates, especially as $L$ 
increases. This needs to be qualified though by recalling the continuity and 
monotonicity criteria which means the available number of satisfactory 
approximants will be less than either total. What the specific number is for 
any exponent can not be determined until explicit expressions for each 
approximant are constructed. 

Therefore we have devised a sifting algorithm to isolate bona fide exponents. 
First, approximants which have a discontinuity anywhere in the range 
$2$~$\leq$~$d$~$\leq$~$6$ are disregarded even if the discontinuity is located 
below three dimensions. While the four and five dimensional estimates from such
a Pad\'{e} may be in the neighbourhood of values from other methods its lack of 
contiguity with the two dimensional boundary condition would bring doubt into 
the reliabilty of the two higher dimension estimates. The second criteria is to
have monotonicity in the same range so that plots of 
${\cal P}_{[p/q]}(\epsilon)$ should be free from turning or inflection points. 
This sift is based on the evidence of lower order behaviour and the assumption 
that the six loop refinement should not lead to major deviations in the overall
structure. The final condition is not unconnected with the previous one in that
when examining the valid plots of all the Pad\'{e}s for an exponent they have a 
similar appearance. Though we note that locating inflection points, which this 
later sift effectively relates to, is also based on  solving for such points 
analytically. Any Pad\'{e} approximant that passes through the sieve is 
retained for the statistical analysis.

{\begin{table}[ht]
\begin{center}
\begin{tabular}{|c|c||c|c|c||c|c|}
\hline
\rule{0pt}{15pt}
Exp & $d$ & $\mu_{4 \frac{}{}}$ & $\mu_{5 \frac{}{}}$ & $\mu_{6 \frac{}{}}$ & $\mu$ & 
$\mu_{\mbox{\footnotesize{wt}} \frac{}{}}$ \\ 
\hline
\rule{0pt}{15pt}
& $3$ & $-~ 0.582343$ & $-~ 0.571817$ & $-~ 0.577538$ & $-~ 0.578(6)$ & $-~ 0.578(5)$ \\
$\hat{\eta}$ & $4$ & $-~ 0.358010$ & $-~ 0.348864$ & $-~ 0.353143$ & $-~ 0.355(6)$ & $-~ 0.354(6)$ \\
& $5$ & $-~ 0.152577$ & $-~ 0.150706$ & $-~ 0.151357$ & $-~ 0.152(1)$ & $-~ 0.152(1)$ \\
\hline
\rule{0pt}{15pt}
& $3\,$ & $0.076041$ & $0.079097$ & $0.074756$ & $0.078(9)$ & $0.078(8)$ \\
$\hat{\sigma}$ & $4$ & $0.259275$ & $0.261320$ & $0.258843$ & $0.260(6)$ & $0.260(5)$ \\
& $5$ & $0.398278$ & $0.398613$ & $0.398328$ & $0.3984(10)$ & $0.3985(8)$ \\
\hline
\rule{0pt}{15pt}
& $3$ & $0.358377$ & $0.358895$ & $0.357898$ & $0.3585(13)$ & $0.3584(12)$ \\
$\hat{\nu}_c$ & $4$ & $0.314629$ & $0.314999$ & $0.314524$ & $0.3147(7)$ & $0.3147(6)$ \\
& $5$ & $0.279627$ & $0.279692$ & $0.279649$ & $0.2796(2)$ & $0.27965(11)$ \\
\hline
\rule{0pt}{15pt}
& $3$ & --------- & $0.594581$ & $0.569373$ & $0.596(20)$ & $0.590(20)$ \\
$\hat{\theta}$ & $4$ & --------- & $0.385131$ & $0.368677$ & $0.390(16)$ & $0.384(16)$ \\
& $5$ & --------- & $0.196856$ & $0.193744$ & $0.200(6)$ & $0.198(6)$ \\
\hline
\rule{0pt}{15pt}
& $3$ & $1.579682$ & $1.634611$ & $1.602104$ & $1.616(1)$ & $1.615(1)$ \\
$\hat{\omega}$ & $4$ & $1.159805$ & $1.208799$ & $1.181551$ & $1.1928(10)$ & $1.1924(10)$ \\
& $5$ & $0.686603$ & $0.701431$ & $0.694545$ & $0.6965(2)$ & $0.6970(2)$ \\
\hline
\end{tabular}
\caption{Exponent estimates for Lee-Yang theory using constrained Pad\'{e}
approximants containing the four, five and six loop averages, the mean and mean
weighted by loop order for $3$, $4$ and $5$ dimensions.}
\label{lyres}
\end{center}
\end{table}}

From the set of approximants extracted using the above criteria we will 
evaluate several measures of the exponents. The main ones are the mean $\mu$ 
and weighted mean $\mu_{\mbox{\footnotesize{wt}}}$ defined by
\begin{equation}
\mu ~=~ \frac{\sum_m {\cal P}_m}{\sum_n 1} ~~~,~~~ 
\mu_{\mbox{\footnotesize{wt}}} ~=~ \frac{\sum_m w_m {\cal P}_m}{\sum_n w_n}
\end{equation}
where the indexing set on the summation symbols corresponds to the set of valid
approximants for each exponent, represented by ${\cal P}_m$, and $w_m$ are 
non-unit weights. Clearly taking the formal limit $w_m$~$\to$~$1$ in
$\mu_{\mbox{\footnotesize{wt}}}$ produces $\mu$ which also clarifies the 
meaning of the denominator in the definition of $\mu$. We have chosen the $w_m$
to be the loop order that the ${\cal P}_m$ originated from based on the 
assumption that as the loop order increases the higher order corrections should
lead to a more accurate estimate. By the same token we define the uncertainty
via the parallel standard deviations given by
\begin{equation}
\varsigma ~=~ 
\sqrt{\frac{\sum_m \left( \mu - {\cal P}_m \right)^2}{\sum_n 1} } ~~~,~~~
\varsigma_{\mbox{\footnotesize{wt}}} ~=~ 
\sqrt{\frac{\sum_m w_m \left( \mu_{\mbox{\footnotesize{wt}}} 
- {\cal P}_m \right)^2}{\sum_n w_n} }
\end{equation}
with the same aim that $\varsigma_{\mbox{\footnotesize{wt}}}$ should produce a
refined uncertainty in relation to $\varsigma$. 

{\begin{table}[H]
\begin{center}
\begin{tabular}{|c|c||c|c|c|c|c|}
\hline
\rule{0pt}{15pt}
Ref \& Method & $d$ & $\hat{\eta}$ & $\hat{\sigma}$ & $\hat{\nu}_c$ & $\hat{\theta}$ & $\hat{\omega}$ \\
\hline
\rule{0pt}{15pt}
& $1$ & $-~ 0.984(24)$ & $-~ 0.498(3)$ & $0.502(3)$ & $1$ & $1.992(12)$ \\
& $2$ & $-~ 0.77(5)$ & $-~ 0.161(8)$ & $0.420(4)$ & $\frac{5}{6}$ & $1.986(19)$ \\
\cite{26} OA & $3$ & $-~ 0.52(13)$ & $0.0877(25)$ & $0.363(8)$ & $0.622(12)$ & $1.72(8)$ \\
& $4$ & $-~ 0.325(7)$ & $0.2648(15)$ & $0.3162(4)$ & $0.412(8)$ & $1.30(3)$ \\
& $5$ & $-~ 0.13(3)$ & $0.402(5)$ & $0.2804(10)$ & $0.205(5)$ & $0.73(2)$ \\
\hline
\rule{0pt}{15pt}
& $3$ & $-~ 0.576(10)$ & $0.076(2)$ & $0.359(1)$ & --------- & --------- \\
\cite{33} $T$ & $4$ & $-~ 0.36(3)$ & $0.258(5)$ & $0.314(2)$ & --------- & --------- \\
& $5$ & $-~ 0.14(5)$ & $0.401(9)$ & $0.280(2)$ & --------- & --------- \\
\hline
\rule{0pt}{15pt}
\cite{34} CB & $3$ & $-~ 0.574$ & $0.076$ & $0.358$ & --------- & --------- \\
& $4$ & $-~ 0.354$ & $0.259$ & $0.314$ & --------- & --------- \\
\hline
\rule{0pt}{15pt}
& $2$ & $-~ 0.798(12)$ & $-~ 0.1664(20)$ & $0.4168(10)$ & --------- & --------- \\
\cite{35} CB & $3$ & $-~ 0.530(5)$ & $0.085(1)$ & $0.3617(4)$ & --------- & --------- \\
& $4$ & $-~ 0.3067(5)$ & $0.2685(1)$ & $0.3171(3)$ & --------- & --------- \\
& $5$ & $-~ 0.090(3)$ & $0.4105(5)$ & $0.2821(1)$ & --------- & --------- \\
\hline
\rule{0pt}{15pt}
& $3$ & $-~ 0.586(29)$ & $0.0742(56)$ & $0.3581(19)$ & --------- & --------- \\
\cite{27} FRG & $4$ & $-~ 0.316(16)$ & $0.2667(32)$ & $0.3167(8)$ & --------- & --------- \\
& $5$ & $-~ 0.126(6)$ & $0.4033(12)$ & $0.2807(2)$ & --------- & --------- \\
\hline
\rule{0pt}{15pt}
\cite{37} FRG & $4$ & $-~ 0.325(3)$ & $0.2648(6)$ & $0.3162(2)$ & --------- & --------- \\
& $5$ & $-~ 0.1344(1)$ & $0.40166(2)$ & $0.28033(1)$ & --------- & --------- \\
\hline
\rule{0pt}{15pt}
& $3$ & $-~ 0.651$ & $0.062$ & $0.354$ & --------- & --------- \\
\cite{36} CB & $4$ & $-~ 0.353$ & $0.259$ & $0.315$ & --------- & --------- \\
& $5$ & $-~ 0.124$ & $0.404$ & $0.2801$ & --------- & --------- \\
\hline
\rule{0pt}{15pt}
& $3$ & $-~ 0.580(7)$ & $0.078(2)$ & $0.359(1)$ & --------- & --------- \\
\cite{13} $\epsilon$ & $4$ & $-~ 0.356(6)$ & $0.2604(14)$ & $0.3151(3)$ & --------- & --------- \\
& $5$ & $-~ 0.1521(13)$ & $0.3984(2)$ & $0.2797(1)$ & --------- & --------- \\
\hline
\rule{0pt}{15pt}
& $3$ & $-~ 0.564$ & $0.078$ & $0.359$ & --------- & --------- \\
\cite{30} CB & $4$ & $-~ 0.346$ & $0.261$ & $0.315$ & --------- & --------- \\
& $5$ & $-~ 0.140$ & $0.399$ & $0.280$ & --------- & --------- \\
\hline
\rule{0pt}{15pt}
& $3$ & $-~ 0.578(5)$ & $0.078(8)$ & $0.3584(12)$ & $0.590(20)$ & $1.615(1)$ \\
This work & $4$ & $-~ 0.354(6)$ & $0.260(5)$ & $0.3147(6)$ & $0.384(16)$ & $1.1924(10)$ \\
& $5$ & $-~ 0.152(1)$ & $0.3985(8)$ & $0.27965(11)$ & $0.198(6)$ & $0.6970(2)$ \\
\hline
\end{tabular}
\caption{Summary of exponents in various dimensions from different methods
derived from scaling laws using either $\hat{\sigma}$ or $\Delta_\phi$ as 
input.}
\label{lyressum}
\end{center}
\end{table}}

We have applied this process to all the exponents in (\ref{lyexp}) by first 
constructing the Pad\'{e} approximants analytically. Using the numerical 
representation to select the relevant approximants we arrive at the results of 
Table \ref{lyres}. For each exponent estimates are recorded in three different 
dimensions. There are several columns of data. Those headed with $\mu_L$ for 
$L$~$=$~$4$, $5$ and $6$ are the averages of the selected approximants at $L$ 
loops only. These are provided as a guide or an indication of the progression 
of the inclusion of higher order contributions. They are not to be regarded as 
bona fide estimates. Returning to an earlier point that one can never be sure 
which approximants will be reliable it turned out that only seven passed the 
sift test for $\hat{\theta}$ which was the lowest number of all five exponents.
In fact no $\hat{\theta}$ four loop ones did which is the reason there are no 
entries for that exponent. The final two columns contain the two means 
calculated from all Pad\'{e}s that survived the sifting process where the 
uncertainties are deduced from the respective $\varsigma$ and 
$\varsigma_{\mbox{\footnotesize{wt}}}$. There are several general observations.
First the uncertainties are invariably smaller as $d$ increases which can be 
seen across all five exponents in $d$~$=$~$5$ with $\hat{\theta}$ being an 
exception. The reason for this is that there were only seven Pad\'{e}s for 
$\hat{\theta}$ with the absence of four loop estimates affecting the 
uncertainty. In places the uncertainties from the weighted measures are tighter
than those from the usual mean. This is partly because more approximants were 
available at five and six loops and their effect can be gauged in the trends 
apparent in $\mu_L$.

In order to place our six loop exponent estimates in context we have compiled a
summary of previous results in Table \ref{lyressum} which appear in 
chronological order. First we have recorded values for the available dimensions
and note that only two articles provide estimates of the correction to scaling
exponents $\hat{\theta}$ and $\hat{\omega}$. Aside from those from \cite{26} 
three dimensional estimates have been extracted from studies on regularized 
spheres and in particular using a recent technique based on so-called fuzzy 
spheres, \cite{30,31,32}. The respective exponents from \cite{30,31} are 
summarized in Table \ref{lyresfuzz}. For Table \ref{lyressum} other techniques 
were employed such as high temperature expansions, ($T$), \cite{33}, conformal 
bootstrap method, (CB), \cite{30,34,35,36}, functional renormalization group, 
(FRG), \cite{27,37}, $\epsilon$ expansions ($\epsilon$), \cite{13,34,35,36}, or
other approaches (OA), \cite{26,30}. In most of these articles the exponent 
that was calculated was invariably either $\hat{\sigma}$ or $\Delta_\phi$ since
it was shown in \cite{16} that there is only one independent exponent 
reflecting the renormalization properties of (\ref{lagphi3}). In \cite{13}
the $\epsilon$ expansion coupled with the scaling laws produced expressions for
each exponent prior to finding estimates. The values recorded from \cite{13} in
Table \ref{lyressum} are those computed from the five loop constrained Pad\'{e}
estimates in the same way that the results of Table \ref{lyres} were arrived 
at. They are included for comparison with our new order rather than the overall 
constrained values of that paper. Whichever of $\hat{\sigma}$ or $\Delta_\phi$
was determined in an article we derived the others using the scaling laws of
(\ref{scalsigma}) or (\ref{scaldelta}) in order to compare with the constrained 
Pad\'{e}s. In this respect methods that used a fixed dimension approach to
estimate exponents were either unable to provide values for dimensions one or 
two, to test whether they tallied with the exact values, or could make 
predictions which were not in good agreement, or the predictions for these low 
dimensions were indeed reliable. This was the case for \cite{26} as noted in 
Table \ref{lyressum}. We included those $d$~$=$~$1$ and $2$ values since they 
illustrate, for example, one side of the debate on whether the corrections to 
scaling are logarithmic or not.

{\begin{table}[ht]
\begin{center}
\begin{tabular}{|c|c||c|c|c|c|c|}
\hline
\rule{0pt}{15pt}
Ref & Method & $\hat{\eta}_\frac{}{}$ & $\hat{\sigma}$ & $\hat{\nu}_c$ & $\hat{\theta}$ & $\hat{\omega}$ \\
\hline
\rule{0pt}{15pt}
\cite{30} & $E$ & $-~ 0.572(4)$ & $0.0768(8)$ & $0.3589(3)$ & $0.579(3)$ & $1.613(6)$ \\
& $Z$ & $-~ 0.5790(32)$ & $0.0774(6)$ & $0.3591(2)$ & --------- & --------- \\
& $X$ & $-~ 0.5698(16)$ & $0.0772(3)$ & $0.3591(2)$ & --------- & --------- \\
\hline
\rule{0pt}{15pt}
\cite{31} & --------- & $-~ 0.42$ & $0.11$ & $0.37$ & $0.63$ & $1.71$ \\
\hline
\end{tabular}
\caption{Estimates of three dimensional exponents using the value of
$\Delta_\phi$ as input to the scaling laws from \cite{30} where $E$, $Z$ and
$X$ denote eigenvalues and the matrix element of fuzzy spheres $Z$ and $X$
respectively, as well as the estimates from \cite{31}.}
\label{lyresfuzz}
\end{center}
\end{table}}

While the majority of exponent estimates arise from more modern techniques,
such as the functional renormalization group and conformal bootstrap which are
continuum field theory based, it is worth discussing our estimates in 
comparison to those. In particular we concentrate on $\hat{\eta}$,
$\hat{\sigma}$ and to a lesser extent on $\hat{\omega}$ as these tend to have 
been directly measured. What is interesting is that the three dimensional 
values of $\hat{\eta}$, aside from \cite{36}, are very much in accord. In four 
dimensions a similar picture is apparent but perhaps with much less overlap of 
uncertainties. In five dimensions the deviation of $\hat{\eta}$ of perturbation
theory from the two continuum techniques is more distinct. This seems peculiar 
in that it might have been expected that the five dimensional perturbative 
results should be more accurate when summing down from the critical dimension. 
The much earlier estimates of \cite{26,33} for $\hat{\eta}$ do not lean towards
either side due to the large uncertainties. The position with $\hat{\sigma}$ is
not dissimilar unsurprisingly since most methods derive an estimate for 
$\hat{\eta}$, or $\Delta_\phi$ in the case of the conformal bootstrap, and 
apply the scaling relations (\ref{scaldelta}). What is reassuring in the 
comparison of our results with other approaches is the consistency of the 
two-sided Pad\'{e} construction indicating that the resummation down from six 
dimensions can indeed capture low dimension properties. Finally the situation 
with our $\hat{\omega}$ values is not as conclusive merely because there 
appears to be only one earlier analysis across all three discrete dimensions, 
\cite{26}. In that instance our three estimates do not overlap with those of 
\cite{26}. However comparing our estimate for $\hat{\omega}$ with the recent 
use of the fuzzy sphere method in three dimensions, \cite{30}, there is good 
agreement. By way of a final comment on resummation, an alternative approach to
potentially improve convergence of the high order $\epsilon$ series is to use 
the Pad\'{e}-Borel method. This involves first replacing the unconstrained 
series itself by a Borel transform. Its integrand is related to the original 
series but with the coefficients of $\epsilon$ weighted by $1/L!$ where $L$ is 
the loop order. The resulting series in the Borel variable is then replaced by 
a Pad\'{e} approximant before evaluating the transform for the three values of 
$\epsilon$ used here. However in following this procedure it is not possible to
constrain the integrand approximant by the low dimensional exponent values. 
Hence the bridge to low dimensions afforded by exact exponent values in one and
two dimensions, and of great benefit to the direct Pad\'{e} approach, cannot be
accessed in the Pad\'{e}-Borel case.

\sect{Percolation exponents.}

This section is devoted to a similar analysis to that of the previous one but
for percolation theory except that we will provide a constrained Pad\'{e} 
analysis for a larger number of exponents. Like the Lee-Yang model the 
underlying continuum quantum field theory governing criticality is a cubic
scalar theory with a critical dimension of six. More specifically it 
corresponds to the replica limit of the $(N+1)$-state Potts model \cite{38}. 
The renormalization group functions for percolation theory were determined to 
five loops in \cite{13,22,23,24,39,40}. More recently their extension was given
to six loops in the {\sc Hyperlog} package of \cite{15}. Unlike the previous 
section {\em two} core renormalization group functions are the key to 
extracting the main scaling dimensions which are the field and mass operator 
anomalous dimensions $\gamma_\phi(g)$ and $\gamma_{\cal O}(g)$ respectively 
with ${\cal O}$~$=$~$\half \phi^2$. The correspondence with the critical 
exponents is
\begin{equation}
\eta ~=~ \gamma_\phi(g^\star) ~~~,~~~
\frac{1}{\nu} ~=~ 2 ~-~ \eta ~+~ \gamma_{\cal O}(g^\star) ~~~,~~~
\omega ~=~ \beta^\prime(g^\star) 
\end{equation}
while the $\beta$-function connects with the correction to scaling exponent
$\omega$. Using the results of \cite{15} we find
\begin{eqnarray}
\eta &=&
-~ \frac{1}{21} \epsilon
- \frac{206}{9261} \epsilon^2
+ \left[
\frac{256}{7203} \zeta_3
- \frac{93619}{8168202}
\right] \epsilon^3
\nonumber \\
&&
+ \left[
\frac{64}{2401} \zeta_4
+ \frac{189376}{9529569} \zeta_3
- \frac{320}{3087} \zeta_5
- \frac{103309103}{14408708328}
\right] \epsilon^4
\nonumber \\
&&
+ \left[
\frac{47344}{3176523} \zeta_4
+ \frac{2337824}{9529569} \zeta_5
+ \frac{77003747}{600362847} \zeta_3
- \frac{187744}{7411887} \zeta_3^2
- \frac{664}{16807} \zeta_7
- \frac{400}{3087} \zeta_6
\right. \nonumber \\
&& \left. ~~~~
-~ \frac{43137745921}{3630994498656}
\right] \epsilon^5
\nonumber \\
&&
+ \left[
\frac{77003747}{800483796} \zeta_4
- \frac{13252084726607}{2135024765209728}
- \frac{350196946003}{1059040062108} \zeta_3
- \frac{325047556}{600362847} \zeta_5
\right. \nonumber \\
&& \left. ~~~~
-~ \frac{7328344}{466948881} \zeta_3^2
- \frac{2079575}{352947} \zeta_7
- \frac{1812032}{7411887} \zeta_3 \zeta_5
- \frac{1159098}{4117715} \zeta_8
- \frac{93872}{2470629} \zeta_3 \zeta_4
+ \frac{37376}{352947} \zeta_3^3
\right. \nonumber \\
&& \left. ~~~~
+~ \frac{361824}{4117715} \zeta_{5,3}
+ \frac{2782048}{453789} \zeta_9
+ \frac{2816440}{9529569} \zeta_6
\right] \epsilon^6 ~+~ O(\epsilon^7) \nonumber \\
\frac{1}{\nu} &=&
2
- \frac{5}{21} \epsilon
- \frac{653}{18522} \epsilon^2
+ \left[
\frac{356}{7203} \zeta_3
- \frac{332009}{32672808}
\right] \epsilon^3
\nonumber \\
&&
+ \left[
\frac{110219}{9529569} \zeta_3
- \frac{3760}{21609} \zeta_5
+ \frac{89}{2401} \zeta_4
- \frac{59591131}{57634833312}
\right] \epsilon^4
\nonumber \\
&&
+ \left[
\frac{298060003}{2401451388} \zeta_3
- \frac{119568216869}{14523977994624}
- \frac{134000}{7411887} \zeta_3^2
- \frac{4700}{21609} \zeta_6
+ \frac{2952}{16807} \zeta_7
+ \frac{110219}{12706092} \zeta_4
\right. \nonumber \\
&& \left. ~~~~
+~ \frac{3242404}{9529569} \zeta_5
\right] \epsilon^5
\nonumber \\
&&
+ \left[
\frac{298060003}{3201935184} \zeta_4
- \frac{9315646748605}{8540099060838912}
- \frac{2028254681339}{4236160248432} \zeta_3
- \frac{375720260}{600362847} \zeta_5
- \frac{17931317}{2117682} \zeta_7
\right. \nonumber \\
&& \left. ~~~~
-~ \frac{12592520}{466948881} \zeta_3^2
- \frac{3902656}{7411887} \zeta_3 \zeta_5
- \frac{67000}{2470629} \zeta_3 \zeta_4
+ \frac{46976}{352947} \zeta_3^3
+ \frac{224544}{4117715} \zeta_{5,3}
\right. \nonumber \\
&& \left. ~~~~
+~ \frac{723022}{4117715} \zeta_8
+ \frac{15623285}{38118276} \zeta_6
+ \frac{24682792}{3176523} \zeta_9
\right] \epsilon^6
~+~ O(\epsilon^7) \nonumber \\
\omega &=&
\epsilon
- \frac{671}{882} \epsilon^2
+ \left[
\frac{372}{343} \zeta_3
+ \frac{40639}{57624}
\right] \epsilon^3
+ \left[
 \frac{279}{343} \zeta_4
- \frac{348539}{151263} \zeta_3
- \frac{1360}{343} \zeta_5
- \frac{317288185}{304946208}
\right] \epsilon^4
\nonumber \\
&&
+ \left[
\frac{11664257531}{800483796} \zeta_3
+ \frac{601352852897}{691617999744}
- \frac{348539}{201684} \zeta_4
- \frac{55596}{2401} \zeta_7
- \frac{1700}{343} \zeta_6
+ \frac{207440}{117649} \zeta_3^2
\right. \nonumber \\
&& \left. ~~~~
+~ \frac{17305178}{453789} \zeta_5
\right] \epsilon^5
\nonumber \\
&&
+ \left[
\frac{43023079}{7411887} \zeta_3^2
- \frac{17253383458933}{201721916592} \zeta_3
- \frac{59563537247}{400241898} \zeta_5
+ \frac{11664257531}{1067311728} \zeta_4
\right. \nonumber \\
&& \left. ~~~~
+~ \frac{33644508241033}{406671383849472}
- \frac{1388669831}{2117682} \zeta_7
- \frac{13215057}{117649} \zeta_8
- \frac{10633120}{352947} \zeta_3 \zeta_5
+ \frac{46720}{2401} \zeta_3^3
\right. \nonumber \\
&& \left. ~~~~
+~ \frac{311160}{117649} \zeta_3 \zeta_4
+ \frac{3495600}{117649} \zeta_{5,3}
+ \frac{85910695}{1815156} \zeta_6
+ \frac{122890160}{151263} \zeta_9
\right] \epsilon^6 ~+~ O(\epsilon^7) 
\end{eqnarray} 
to six loops. These three exponents then appear in the definition of scaling
dimensions through the scaling laws
\begin{eqnarray}
\alpha &=& 2 ~-~ d \nu ~~,~~
\beta ~=~ \frac{1}{2} [ d - 2 + \eta ] \nu ~~,~~
\gamma ~=~ [ 2 - \eta ] \nu ~~,~~
\delta ~=~ \frac{[ d + 2 - \eta ]}{[ d - 2 + \eta ]} ~~,~~
\nonumber \\
\sigma &=& \frac{2}{[ d + 2 - \eta ] \nu} ~~,~~
\tau ~=~ 1 ~+~ \frac{2d}{[ d + 2 - \eta ]} ~~,~~
d_f ~=~ \frac{1}{2} [ d + 2 - \eta ] ~~,~~
\Omega ~=~ \frac{2 \omega}{[ d + 2 - \eta ]} ~~~~~~~~
\label{percscal}
\end{eqnarray}
where $\Omega$ is a second correction to scaling exponent. In Lee-Yang theory
it equates to $\hat{\theta}$ while the fractal dimension, $d_f$, is equivalent
to $1/\hat{\nu}_c$. We note that the fractal dimension was denoted by $D$ in 
\cite{41}.

Using these scaling laws we deduce
\begin{eqnarray}
\alpha &=&
-~ 1
+ \frac{1}{7} \epsilon
- \frac{443}{12348} \epsilon^2
+ \left[
\frac{178}{2401} \zeta_3
- \frac{370187}{21781872}
\right] \epsilon^3
\nonumber \\
&&
+ \left[
\frac{267}{4802} \zeta_4
+ \frac{143861}{6353046} \zeta_3
- \frac{1880}{7203} \zeta_5
- \frac{133741081}{38423222208}
\right] \epsilon^4
\nonumber \\
&&
+ \left[
\frac{143861}{8470728} \zeta_4
+ \frac{1561982}{3176523} \zeta_5
+ \frac{304564789}{1600967592} \zeta_3
- \frac{67000}{2470629} \zeta_3^2
- \frac{2350}{7203} \zeta_6
+ \frac{4428}{16807} \zeta_7
\right. \nonumber \\
&& \left. ~~~~
-~ \frac{127531100903}{9682651996416}
\right] \epsilon^5
\nonumber \\
&&
+ \left[
\frac{23488}{117649} \zeta_3^3
- \frac{1985370060917}{2824106832288} \zeta_3
- \frac{1760162509991}{632599930432512}
- \frac{182554231}{200120949} \zeta_5
- \frac{17904749}{1411788} \zeta_7
\right. \nonumber \\
&& \left. ~~~~
-~ \frac{6882916}{155649627} \zeta_3^2
- \frac{1951328}{2470629} \zeta_3 \zeta_5
- \frac{33500}{823543} \zeta_3 \zeta_4
+ \frac{336816}{4117715} \zeta_{5,3}
+ \frac{1084533}{4117715} \zeta_8
\right. \nonumber \\
&& \left. ~~~~
+~ \frac{12341396}{1058841} \zeta_9
+ \frac{15031085}{25412184} \zeta_6
+ \frac{304564789}{2134623456} \zeta_4
\right] \epsilon^6
~+~ O(\epsilon^7) \nonumber \\
\beta &=&
1
- \frac{1}{7} \epsilon
- \frac{61}{12348} \epsilon^2
- \left[
\frac{19405}{21781872}
+ \frac{38}{2401} \zeta_3
\right] \epsilon^3
\nonumber \\
&&
+ \left[
\frac{440}{7203} \zeta_5
+ \frac{5281}{6353046} \zeta_3
- \frac{57}{4802} \zeta_4
- \frac{84318803}{38423222208}
\right] \epsilon^4
\nonumber \\
&&
+ \left[
\frac{7457829179}{9682651996416}
- \frac{46787653}{1600967592} \zeta_3
- \frac{13406}{117649} \zeta_5
- \frac{1642}{16807} \zeta_7
+ \frac{550}{7203} \zeta_6
+ \frac{5281}{8470728} \zeta_4
\right. \nonumber \\
&& \left. ~~~~
+~ \frac{6688}{2470629} \zeta_3^2
\right] \epsilon^5
\nonumber \\
&&
+ \left[
\frac{457789037921}{2824106832288} \zeta_3
- \frac{2938640361995}{1897799791297536}
- \frac{46787653}{2134623456} \zeta_4
- \frac{7472812}{3176523} \zeta_9
- \frac{1302571}{8235430} \zeta_8
\right. \nonumber \\
&& \left. ~~~~
-~ \frac{129405}{941192} \zeta_6
- \frac{21816}{4117715} \zeta_{5,3}
- \frac{14144}{352947} \zeta_3^3
+ \frac{3344}{823543} \zeta_3 \zeta_4
+ \frac{166480}{823543} \zeta_3 \zeta_5
+ \frac{417731}{151263} \zeta_7
\right. \nonumber \\
&& \left. ~~~~
+~ \frac{1398094}{155649627} \zeta_3^2
+ \frac{12603706}{66706983} \zeta_5
\right] \epsilon^6
~+~ O(\epsilon^7) \nonumber \\
\delta &=&
2
+ \frac{2}{7} \epsilon
+ \frac{565}{6174} \epsilon^2
+ \left[
\frac{371953}{10890936}
- \frac{64}{2401} \zeta_3
\right] \epsilon^3
\nonumber \\
&&
+ \left[
\frac{300656141}{19211611104}
- \frac{77584}{3176523} \zeta_3
- \frac{48}{2401} \zeta_4
+ \frac{80}{1029} \zeta_5
\right] \epsilon^4
\nonumber \\
&&
+ \left[
\frac{67649411155}{4841325998208}
- \frac{84030083}{800483796} \zeta_3
- \frac{496256}{3176523} \zeta_5
- \frac{19396}{1058841} \zeta_4
+ \frac{100}{1029} \zeta_6
+ \frac{498}{16807} \zeta_7
\right. \nonumber \\
&& \left. ~~~~
+~ \frac{46936}{2470629} \zeta_3^2
\right] \epsilon^5
\nonumber \\
&&
+ \left[
\frac{297068306929}{1412053416144} \zeta_3
+ \frac{9026360188351}{948899895648768}
- \frac{84030083}{1067311728} \zeta_4
- \frac{695512}{151263} \zeta_9
- \frac{593860}{3176523} \zeta_6
\right. \nonumber \\
&& \left. ~~~~
-~ \frac{271368}{4117715} \zeta_{5,3}
- \frac{9344}{117649} \zeta_3^3
+ \frac{23468}{823543} \zeta_3 \zeta_4
+ \frac{453008}{2470629} \zeta_3 \zeta_5
+ \frac{1738647}{8235430} \zeta_8
+ \frac{2084555}{470596} \zeta_7
\right. \nonumber \\
&& \left. ~~~~
+~ \frac{2925010}{155649627} \zeta_3^2
+ \frac{70128049}{200120949} \zeta_5
\right] \epsilon^6
~+~ O(\epsilon^7) \nonumber \\
d_f &=&
4
- \frac{10}{21} \epsilon
+ \frac{103}{9261} \epsilon^2
+ \left[
\frac{93619}{16336404}
- \frac{128}{7203} \zeta_3
\right] \epsilon^3
\nonumber \\
&&
+ \left[
\frac{103309103}{28817416656}
- \frac{94688}{9529569} \zeta_3
- \frac{32}{2401} \zeta_4
+ \frac{160}{3087} \zeta_5
\right] \epsilon^4
\nonumber \\
&&
+ \left[
\frac{43137745921}{7261988997312}
- \frac{77003747}{1200725694} \zeta_3
- \frac{1168912}{9529569} \zeta_5
- \frac{23672}{3176523} \zeta_4
+ \frac{200}{3087} \zeta_6
+ \frac{332}{16807} \zeta_7
\right. \nonumber \\
&& \left. ~~~~
+~ \frac{93872}{7411887} \zeta_3^2
\right] \epsilon^5
\nonumber \\
&&
+ \left[
\frac{350196946003}{2118080124216} \zeta_3
+ \frac{13252084726607}{4270049530419456}
- \frac{77003747}{1600967592} \zeta_4
- \frac{1408220}{9529569} \zeta_6
- \frac{1391024}{453789} \zeta_9
\right. \nonumber \\
&& \left. ~~~~
-~ \frac{180912}{4117715} \zeta_{5,3}
- \frac{18688}{352947} \zeta_3^3
+ \frac{46936}{2470629} \zeta_3 \zeta_4
+ \frac{579549}{4117715} \zeta_8
+ \frac{906016}{7411887} \zeta_3 \zeta_5
+ \frac{2079575}{705894} \zeta_7
\right. \nonumber \\
&& \left. ~~~~
+~ \frac{3664172}{466948881} \zeta_3^2
+ \frac{162523778}{600362847} \zeta_5
\right] \epsilon^6
~+~ O(\epsilon^7) \nonumber \\
\gamma &=&
1
+ \frac{1}{7} \epsilon
+ \frac{565}{12348} \epsilon^2
+ \left[
\frac{408997}{21781872}
- \frac{102}{2401} \zeta_3
\right] \epsilon^3
\nonumber \\
&&
+ \left[
\frac{302378687}{38423222208}
- \frac{154423}{6353046} \zeta_3
- \frac{153}{4802} \zeta_4
+ \frac{1000}{7203} \zeta_5
\right] \epsilon^4
\nonumber \\
&&
+ \left[
\frac{112615442545}{9682651996416}
- \frac{210989483}{1600967592} \zeta_3
- \frac{838058}{3176523} \zeta_5
- \frac{154423}{8470728} \zeta_4
- \frac{1144}{16807} \zeta_7
+ \frac{1250}{7203} \zeta_6
\right. \nonumber \\
&& \left. ~~~~
+~ \frac{53624}{2470629} \zeta_3^2
\right] \epsilon^5
\nonumber \\
&&
+ \left[
\frac{1069791985075}{2824106832288} \zeta_3
+ \frac{11157768253963}{1897799791297536}
- \frac{210989483}{2134623456} \zeta_4
- \frac{22078564}{3176523} \zeta_9
- \frac{8043215}{25412184} \zeta_6
\right. \nonumber \\
&& \left. ~~~~
-~ \frac{293184}{4117715} \zeta_{5,3}
- \frac{42176}{352947} \zeta_3^3
+ \frac{26812}{823543} \zeta_3 \zeta_4
+ \frac{136064}{352947} \zeta_3 \zeta_5
+ \frac{218038}{4117715} \zeta_8
+ \frac{4086728}{155649627} \zeta_3^2
\right. \nonumber \\
&& \left. ~~~~
+~ \frac{30321311}{4235364} \zeta_7
+ \frac{106931995}{200120949} \zeta_5
\right] \epsilon^6
~+~ O(\epsilon^7) \nonumber \\
\nu &=&
\frac{1}{2}
+ \frac{5}{84} \epsilon
+ \frac{589}{37044} \epsilon^2
+ \left[
\frac{716519}{130691232}
- \frac{89}{7203} \zeta_3
\right] \epsilon^3
\nonumber \\
&&
+ \left[
\frac{344397667}{230539333248}
- \frac{222359}{38118276} \zeta_3
- \frac{89}{9604} \zeta_4
+ \frac{940}{21609} \zeta_5
\right] \epsilon^4
\nonumber \\
&&
+ \left[
\frac{33500}{7411887} \zeta_3^2
+ \frac{141995802917}{58095911978496}
- \frac{313903867}{9605805552} \zeta_3
- \frac{711901}{9529569} \zeta_5
- \frac{222359}{50824368} \zeta_4
- \frac{738}{16807} \zeta_7
\right. \nonumber \\
&& \left. ~~~~
+~ \frac{1175}{21609} \zeta_6
\right] \epsilon^5
\nonumber \\
&&
+ \left[
\frac{1893082324019}{16944640993728} \zeta_3
+ \frac{29757051275785}{34160396243355648}
- \frac{313903867}{12807740736} \zeta_4
- \frac{13649285}{152473104} \zeta_6
\right. \nonumber \\
&& \left. ~~~~
-~ \frac{6170698}{3176523} \zeta_9
- \frac{361511}{8235430} \zeta_8
- \frac{56136}{4117715} \zeta_{5,3}
- \frac{11744}{352947} \zeta_3^3
+ \frac{16750}{2470629} \zeta_3 \zeta_4
+ \frac{975664}{7411887} \zeta_3 \zeta_5
\right. \nonumber \\
&& \left. ~~~~
+~ \frac{3793208}{466948881} \zeta_3^2
+ \frac{17842757}{8470728} \zeta_7
+ \frac{83802155}{600362847} \zeta_5
\right] \epsilon^6
~+~ O(\epsilon^7) \nonumber \\
\sigma &=&
\frac{1}{2}
- \frac{1}{98} \epsilon^2
+ \left[
\frac{5}{343} \zeta_3
- \frac{773}{172872}
\right] \epsilon^3
+ \left[
\frac{3551}{605052} \zeta_3
+ \frac{15}{1372} \zeta_4
- \frac{120}{2401} \zeta_5
- \frac{246103}{203297472}
\right] \epsilon^4
\nonumber \\
&&
+ \left[
\frac{1763659}{44471322} \zeta_3
- \frac{149476871}{51230962944}
- \frac{718}{117649} \zeta_3^2
- \frac{150}{2401} \zeta_6
+ \frac{199}{4802} \zeta_7
+ \frac{3551}{806736} \zeta_4
\right. \nonumber \\
&& \left. ~~~~
+~ \frac{300005}{3176523} \zeta_5
\right] \epsilon^5
\nonumber \\
&&
+ \left[
\frac{7387844}{3176523} \zeta_9
- \frac{18252766321}{134481277728} \zeta_3
- \frac{4189853447}{4236160248432}
- \frac{143110153}{800483796} \zeta_5
- \frac{42017779}{16941456} \zeta_7
\right. \nonumber \\
&& \left. ~~~~
-~ \frac{362972}{2470629} \zeta_3 \zeta_5
- \frac{124291}{14823774} \zeta_3^2
- \frac{1077}{117649} \zeta_3 \zeta_4
+ \frac{2250}{117649} \zeta_{5,3}
+ \frac{14080}{352947} \zeta_3^3
+ \frac{24757}{941192} \zeta_8
\right. \nonumber \\
&& \left. ~~~~
+~ \frac{1763659}{59295096} \zeta_4
+ \frac{5768575}{50824368} \zeta_6
\right] \epsilon^6
~+~ O(\epsilon^7) \nonumber \\
\tau &=&
\frac{5}{2}
- \frac{1}{14} \epsilon
- \frac{313}{24696} \epsilon^2
+ \left[
\frac{16}{2401} \zeta_3
- \frac{150697}{43563744}
\right] \epsilon^3
\nonumber \\
&&
+ \left[
\frac{13348}{3176523} \zeta_3
+ \frac{12}{2401} \zeta_4
- \frac{20}{1029} \zeta_5
- \frac{124383401}{76846444416}
\right] \epsilon^4
\nonumber \\
&&
+ \left[
\frac{77797763}{3201935184} \zeta_3
- \frac{45091644259}{19365303992832}
- \frac{11734}{2470629} \zeta_3^2
- \frac{249}{33614} \zeta_7
- \frac{25}{1029} \zeta_6
+ \frac{3337}{1058841} \zeta_4
\right. \nonumber \\
&& \left. ~~~~
+~ \frac{141704}{3176523} \zeta_5
\right] \epsilon^5
\nonumber \\
&&
+ \left[
\frac{173878}{151263} \zeta_9
- \frac{1662624619001}{1265199860865024}
- \frac{340712273113}{5648213664576} \zeta_3
- \frac{78589417}{800483796} \zeta_5
- \frac{2080571}{1882384} \zeta_7
\right. \nonumber \\
&& \left. ~~~~
-~ \frac{1738647}{32941720} \zeta_8
- \frac{1012433}{311299254} \zeta_3^2
- \frac{113252}{2470629} \zeta_3 \zeta_5
- \frac{5867}{823543} \zeta_3 \zeta_4
+ \frac{2336}{117649} \zeta_3^3
\right. \nonumber \\
&& \left. ~~~~
+~ \frac{67842}{4117715} \zeta_{5,3}
+ \frac{170515}{3176523} \zeta_6
+ \frac{77797763}{4269246912} \zeta_4
\right] \epsilon^6
~+~ O(\epsilon^7) \nonumber \\
\Omega &=&
\frac{1}{4} \epsilon
- \frac{283}{1764} \epsilon^2
+ \left[
\frac{93}{343} \zeta_3
+ \frac{324689}{2074464}
\right] \epsilon^3
+ \left[
\frac{279}{1372} \zeta_4
- \frac{328337}{605052} \zeta_3
- \frac{340}{343} \zeta_5
- \frac{294452729}{1219784832}
\right] \epsilon^4
\nonumber \\
&&
+ \left[
\frac{11454697649}{3201935184} \zeta_3
+ \frac{520662674885}{2766471998976}
- \frac{328337}{806736} \zeta_4
- \frac{13899}{2401} \zeta_7
- \frac{425}{343} \zeta_6
+ \frac{51860}{117649} \zeta_3^2
\right. \nonumber \\
&& \left. ~~~~
+~ \frac{8542549}{907578} \zeta_5
\right] \epsilon^5
\nonumber \\
&&
+ \left[
\frac{84810295}{7260624} \zeta_6
- \frac{16905364832761}{806887666368} \zeta_3
- \frac{57749587781}{1600967592} \zeta_5
+ \frac{11454697649}{4269246912} \zeta_4
\right. \nonumber \\
&& \left. ~~~~
+~ \frac{70447924854521}{1626685535397888}
- \frac{1394517869}{8470728} \zeta_7
- \frac{13215057}{470596} \zeta_8
- \frac{2658280}{352947} \zeta_3 \zeta_5
+ \frac{11680}{2401} \zeta_3^3
\right. \nonumber \\
&& \left. ~~~~
+~ \frac{77790}{117649} \zeta_3 \zeta_4
+ \frac{873900}{117649} \zeta_{5,3}
+ \frac{30722540}{151263} \zeta_9
+ \frac{44591123}{29647548} \zeta_3^2
\right] \epsilon^6 ~+~ O(\epsilon^7) 
\end{eqnarray}
for the remaining exponents. We have included the $\epsilon$ expansion of $\nu$
here since we constucted approximants for both it and $1/\nu$ following the 
approach of \cite{13,24}. Numerically evaluating the expressions gives
\begin{eqnarray}
\alpha &=& -~ 1.000000 
+ 0.142857 \epsilon 
- 0.035876 \epsilon^2 
+ 0.072120 \epsilon^3 
- 0.186722 \epsilon^4 
+ 0.638338 \epsilon^5 
\nonumber \\
&&
-~ 2.633771 \epsilon^6 ~+~ O(\epsilon^7) \nonumber \\
\beta &=& 1.000000 
- 0.142857 \epsilon 
- 0.004940 \epsilon^2 
- 0.019915 \epsilon^3 
+ 0.049299 \epsilon^4 
- 0.168762 \epsilon^5 
\nonumber \\
&&
+~ 0.694682 \epsilon^6 ~+~ O(\epsilon^7) \nonumber \\
\delta &=& 2.000000 
+ 0.285714 \epsilon 
+ 0.091513 \epsilon^2 
+ 0.002111 \epsilon^3 
+ 0.045269 \epsilon^4 
- 0.137837 \epsilon^5 
\nonumber \\
&&
+~ 0.574005 \epsilon^6 ~+~ O(\epsilon^7) \nonumber \\
d_f &=& 4.000000 
- 0.476190 \epsilon 
+ 0.011122 \epsilon^2 
- 0.015630 \epsilon^3 
+ 0.030960 \epsilon^4 
- 0.102275 \epsilon^5 
\nonumber \\
&&
+~ 0.415376 \epsilon^6 ~+~ O(\epsilon^7) \nonumber \\
\eta &=& 
-~ 0.047619 \epsilon 
- 0.022244 \epsilon^2 
+ 0.031261 \epsilon^3 
- 0.061921 \epsilon^4 
+ 0.204551 \epsilon^5 
\nonumber \\
&&
-~ 0.830752 \epsilon^6 ~+~ O(\epsilon^7) \nonumber \\
\gamma &=& 1.000000 
+ 0.142857 \epsilon 
+ 0.045756 \epsilon^2 
- 0.032289 \epsilon^3 
+ 0.088124 \epsilon^4 
- 0.300814 \epsilon^5 
\nonumber \\
&&
+~ 1.244406 \epsilon^6 ~+~ O(\epsilon^7) \nonumber \\
\nu &=& 0.500000 
+ 0.059524 \epsilon 
+ 0.015900 \epsilon^2 
- 0.009370 \epsilon^3 
+ 0.029559 \epsilon^4 
- 0.101463 \epsilon^5 
\nonumber \\
&&
+~ 0.422051 \epsilon^6 ~+~ O(\epsilon^7) \nonumber \\
\frac{1}{\nu} &=& 2.000000 
- 0.238095 \epsilon 
- 0.035255 \epsilon^2 
+ 0.049249 \epsilon^3 
- 0.127439 \epsilon^4 
+ 0.432873 \epsilon^5 
\nonumber \\
&&
-~ 1.780994 \epsilon^6 ~+~ O(\epsilon^7) \nonumber \\
\sigma &=& 0.500000 
- 0.010204 \epsilon^2 
+ 0.013051 \epsilon^3 
- 0.034148 \epsilon^4 
+ 0.116861 \epsilon^5 
\nonumber \\
&&
-~ 0.483033 \epsilon^6 ~+~ O(\epsilon^7) \nonumber \\
\tau &=& 2.500000 
- 0.071429 \epsilon 
- 0.012674 \epsilon^2 
+ 0.004551 \epsilon^3 
- 0.011312 \epsilon^4 
+ 0.037497 \epsilon^5 
\nonumber \\
&&
-~ 0.152981 \epsilon^6 ~+~ O(\epsilon^7) \nonumber \\
\omega &=& \epsilon 
- 0.760771 \epsilon^2 
+ 2.008933 \epsilon^3 
- 7.041302 \epsilon^4 
+ 30.214779 \epsilon^5 
\nonumber \\
&&
-~ 147.820814 \epsilon^6 ~+~ O(\epsilon^7) \nonumber \\
\Omega &=& 0.250000 \epsilon 
- 0.160431 \epsilon^2 
+ 0.482439 \epsilon^3 
- 1.701469 \epsilon^4 
+ 7.347235 \epsilon^5 
\nonumber \\
&&
-~ 36.066283 \epsilon^6 ~+~ O(\epsilon^7) 
\end{eqnarray}
where we note that $\alpha$, $d_f$, $\sigma$, $\omega$ and $\Omega$ are
alternating series to six loops. Equally the general trend in the magnitude of
the coefficient of the new order is similar to the Lee-Yang model since it is
invariably the case that it is an order of magnitude larger than the five loop 
one. However in the analysis of the previous section it was apparent that 
despite this the constrained Pad\'{e} approximants could produce exponent 
estimates that were in close proximity to values from other methods. Therefore 
the expectation is that a parallel analysis will produce percolation exponents 
that are commensurate with other techniques. We have followed the same process 
of constructing the constrained approximants for the $\epsilon$ expansion of 
each exponent with one difference. That is that only one constraint is 
available for a low dimension with the two dimensional critical exponents 
already determined from a minimal conformal field theory with zero central 
charge in \cite{25}. We have recorded these for reference in Table 
\ref{percd26}. In previous work, \cite{24}, a different two dimensional value 
of $\omega$ was used to set up the constrained Pad\'{e} which was $2$ and not 
$\frac{3}{2}$ which is employed here. This is based on the arguments given in 
\cite{41}.

{\begin{table}[hb]
\begin{center}
\begin{tabular}{|c||c|c|c|c|c|c|c|c|c|c|c|}
\hline
\rule{0pt}{15pt}
$d$ & $\alpha_{\frac{}{}}$ & $\beta$ & $\delta$ & $d_{f \frac{}{}}$ & $\eta$ & $\gamma$ & $\nu$ & $\sigma$ & $\tau$ & $\omega$ & $\Omega$ \\
\hline
\rule{0pt}{15pt}
$2$ &
$-~ \frac{2}{3}_{\frac{}{}}$ &
$\frac{5}{36}$ &
$\frac{91}{5}$ &
$\frac{91}{48}$ &
$\frac{5}{24}$ &
$\frac{43}{18}$ &
$\frac{4}{3}$ &
$\frac{36}{91}$ &
$\frac{187}{91}$ &
$\frac{3}{2}$ &
$\frac{72}{91}$ \\
\rule{0pt}{15pt}
$6$ &
$-~ 1_{\frac{}{}}$ &
$1_{\frac{}{}}$ &
$2_{\frac{}{}}$ &
$4_{\frac{}{}}$ &
$0$ &
$1_{\frac{}{}}$ &
$\frac{1}{2}_{\frac{}{}}$ &
$\frac{1}{2}_{\frac{}{}}$ &
$\frac{5}{2}_{\frac{}{}}$ &
$0$ &
$0$ \\
\hline
\end{tabular}
\caption{Values of percolation exponents in two and six dimensions.}
\label{percd26}
\end{center}
\end{table}}

Recalling that to six loops there is a total of $25$ potential approximants, we
have constructed the two-sided Pad\'{e} approximants with one constraint for 
each exponent. Of the $25$ we have removed those $d$-dimensional expressions 
which fail the same the sifting test used for Lee-Yang theory. In the case of
$\eta$ no approximants survived. For $\alpha$ and $\Omega$ only nine Pad\'{e}s
could be used while for $\omega$ it was eight which was the smallest number for
this set of exponents. Indeed this aspect of the analysis is reflected in Table
\ref{percres} which records our results in a parallel way to Table \ref{lyres}.
For instance for these three exponents there was no five loop approximant
for $\alpha$ and only one for each of $\omega$ and $\Omega$ which could be used
for the means of the final two columns of Table \ref{percres}. For $\alpha$ 
there was a degree of stability indicated by the individual four and six loop 
averages and reflected in the uncertainties. For the two correction to scaling 
exponents $\Omega$ has a similar stability over the three higher loop means. 
For $\omega$ this is not the case with the individual loop averages increasing 
at each loop order. For the remaining exponents between $12$ and $18$ Pad\'{e}
approximants survived the sieving process with $\delta$, $\sigma$ and $\tau$ 
being at the upper end. 

{\begin{table}[H]
\begin{center}
\begin{tabular}{|c|c||c|c|c||c|c|}
\hline
\rule{0pt}{15pt}
Exp & $d$ & $\mu_{4 \frac{}{}}$ & $\mu_{5 \frac{}{}}$ & $\mu_{6 \frac{}{}}$ & $\mu$ & $\mu_{{\mbox{\footnotesize{wt}} \frac{}{}}}$ \\ 
\hline
\rule{0pt}{15pt}
& $3$ & $-~ 0.701254$ & --------- & $-~ 0.690435$ & $-~ 0.71(2)$ & $-~ 0.704(14)$ \\
$\alpha$
& $4$ & $-~ 0.772721$ & --------- & $-~ 0.763521$ & $-~ 0.78(1)$ & $-~ 0.775(12)$ \\
& $5$ & $-~ 0.873788$ & --------- & $-~ 0.871315$ & $-~ 0.877(5)$ & $-~ 0.875(4)$ \\
\hline
\rule{0pt}{15pt}
& $3$ & $0.428833$ & $0.414161$ & $0.424477$ & $0.43(2)$ & $0.425(15)$ \\
$\beta$ 
& $4$ & $0.659279$ & $0.651654$ & $0.656490$ & $0.66(1)$ & $0.657(8)$ \\
& $5$ & $0.845736$ & $0.844668$ & $0.845156$ & $0.846(2)$ & $0.845(2)$ \\
\hline
\rule{0pt}{15pt}
& $3$ & $5.207390$ & $5.073557$ & $5.187514$ & $5.1(2)$ & $5.1(2)$ \\
$\delta$
& $4$ & $3.185867$ & $3.159635$ & $3.180578$ & $3.17(4)$ & $3.17(3)$ \\
& $5$ & $2.396300$ & $2.394183$ & $2.395631$ & $2.395(4)$ & $2.395(3)$ \\
\hline
\rule{0pt}{15pt}
& $3$ & $5.207142$ & $5.072673$ & $5.187413$ & $5.16(10)$ & $5.15(9)$ \\
$\delta^\dagger$ & $4$ & $3.185852$ & $3.159604$ & $3.180573$ & $3.18(3)$ & $3.17(2)$ \\
& $5$ & $2.396300$ & $2.394183$ & $2.395631$ & $2.396(3)$ & $2.395(2)$ \\
\hline
\rule{0pt}{15pt}
& $3$ & $2.512775$ & $2.500241$ & $2.525863$ & $2.51(2)$ & $2.51(2)$ \\
$d_f$
& $4$ & $3.042408$ & $3.036236$ & $3.046923$ & $3.040(9)$ & $3.041(8)$ \\
& $5$ & $3.527550$ & $3.526712$ & $3.527823$ & $3.527(2)$ & $3.527(1)$ \\
\hline
\rule{0pt}{15pt}
& $3$ & $1.845780$ & $1.771066$ & $1.827194$ & $1.81(5)$ & $1.81(4)$ \\
$\gamma$
& $4$ & $1.453871$ & $1.422552$ & $1.443809$ & $1.44(2)$ & $1.437(18)$ \\
& $5$ & $1.182134$ & $1.178250$ & $1.180328$ & $1.180(4)$ & $1.180(3)$ \\
\hline
\rule{0pt}{15pt}
& $3$ & $0.909753$ & $0.886821$ & $0.899178$ & $0.90(2)$ & $0.90(2)$ \\
$\nu$
& $4$ & $0.696981$ & $0.687393$ & $0.691739$ & $0.694(9)$ & $0.693(8)$ \\
& $5$ & $0.575161$ & $0.573989$ & $0.574299$ & $0.575(1)$ & $0.575(1)$ \\
\hline
\rule{0pt}{15pt}
& $3$ & $0.896445$ & $0.878554$ & $0.887482$ & $0.89(1)$ & $0.889(15)$ \\
$\nu^\dagger$
& $4$ & $0.691889$ & $0.686013$ & $0.688240$ & $0.690(5)$ & $0.689(5)$ \\
& $5$ & $0.574628$ & $0.574023$ & $0.574080$ & $0.5745(9)$ & $0.5743(7)$ \\
\hline
\rule{0pt}{15pt}
& $3$ & $0.441521$ & $0.456470$ & $0.445084$ & $0.444(9)$ & $0.446(8)$ \\
$\sigma$
& $4$ & $0.474008$ & $0.482269$ & $0.476428$ & $0.475(6)$ & $0.476(5)$ \\
& $5$ & $0.493211$ & $0.494452$ & $0.493727$ & $0.4931(13)$ & $0.4934(11)$ \\
\hline
\rule{0pt}{15pt}
& $3$ & $2.192358$ & $2.195820$ & $2.192879$ & $2.195(5)$ & $2.194(4)$ \\
$\tau$
& $4$ & $2.314033$ & $2.316064$ & $2.314442$ & $2.315(3)$ & $2.315(2)$ \\
& $5$ & $2.417326$ & $2.417649$ & $2.417428$ & $2.4174(7)$ & $2.4175(5)$ \\
\hline
\rule{0pt}{15pt}
& $3$ & $1.362005$ & $1.451995^\ast$ & $1.521970$ & $1.41(11)$ & $1.44(10)$ \\
$\omega$
& $4$ & $1.115530$ & $1.195616^\ast$ & $1.248740$ & $1.02(16)$ & $1.18(9)$ \\
& $5$ & $0.702260$ & $0.727000^\ast$ & $0.740577$ & $0.63(9)$ & $0.72(3)$ \\
\hline
\rule{0pt}{15pt}
& $3$ & $0.597626$ & $0.596757^\ast$ & $0.607541$ & $0.601(6)$ & $0.602(7)$ \\
$\Omega$
& $4$ & $0.403609$ & $0.402615^\ast$ & $0.410698$ & $0.406(4)$ & $0.407(5)$ \\
& $5$ & $0.208204$ & $0.207715^\ast$ & $0.209611$ & $0.2088(11)$ & $0.2089(11)$ \\
\hline
\end{tabular}
\caption{Exponent estimates for percolation using constrained Pad\'{e}
approximants containing the four, five and six loop averages, the mean and mean
weighted by loop order for $3$, $4$ and $5$ dimensions. The estimates for
$\delta^\dagger$ and $\nu^\dagger$ were produced from approximants for 
$1/\delta$ and $1/\nu$ respectively. Entries marked with ${}^\ast$ indicate 
only one valid approximant was available.}
\label{percres}
\end{center}
\end{table}}

{\begin{table}[ht]
\begin{center}
\begin{tabular}{|c|c||c|c|c||c|c|}
\hline
\rule{0pt}{15pt}
Exp & $d$ & $\mu_{4 \frac{}{}}$ & $\mu_{5 \frac{}{}}$ & $\mu_{6 \frac{}{}}$ & $\mu$ & $\mu_{{\mbox{\footnotesize{wt}} \frac{}{}}}$ \\ 
\hline
\rule{0pt}{15pt}
& $3$ & $5.207390$ & $5.073557$ & $5.187514$ & $5.15(14)$ & $5.14(12)$ \\
$\delta$ & $4$ & $3.185867$ & $3.159635$ & $3.180578$ & $3.17(4)$ & $3.17(3)$ \\
& $5$ & $2.396300$ & $2.394183$ & $2.395631$ & $2.396(3)$ & $2.395(3)$ \\
\hline
\rule{0pt}{15pt}
& $3$ & $0.904049$ & $0.883522$ & $0.894166$ & $0.90(2)$ & $0.895(20)$ \\
$\nu$ & $4$ & $0.694799$ & $0.686835$ & $0.690240$ & $0.693(7)$ & $0.691(8)$ \\
& $5$ & $0.574932$ & $0.574004$ & $0.574205$ & $0.575(1)$ & $0.5745(10)$ \\
\hline
\end{tabular}
\caption{Estimates for $\delta$ and $\nu$ from combining data from all 
available approximants for each exponent.}
\label{perccombres}
\end{center}
\end{table}}

For the former there is a marked difference in the estimates across the three 
dimensions in that the central value is less reliable as a guide to behaviour 
as $d$ decreases. This lies in the fact that the difference in the endpoint 
boundaries at two and six dimensions, respectively $18.5$ and $2$, is by far 
the largest range of the exponent set as can be seen in Table \ref{percd26}. So 
it would be difficult to produce the level of uncertainty of the five 
dimensional estimate in the three dimensions. To try to address this we 
followed the same strategy used in \cite{13} which was to construct two-sided 
Pad\'{e}s for $1/\delta$ in $d$-dimensions and deduce estimates for $\delta$ 
from the set of valid approximants. In this approach the magnitude of 
$1/\delta$ only varies by around $0.2$ between the boundary dimensions in 
contrast to $16.5$ for $\delta$. The associated results for this are recorded 
against the entry denoted as $\delta^\dagger$ in Table \ref{percres}. While the
respective central values are similar to the direct approach for $\delta$ the
uncertainties are smaller. The situation with $\sigma$ and $\tau$ is cleaner in
that with the endpoint differences being two orders of magnitude less than that
of $\delta$ the central values are settled judging by the trend of the four, 
five and six loop means. This results in significantly smaller uncertainties in
a dimensional comparison with $\delta$ with possibly a more accurate three
dimensional estimate.

{\begin{table}[hb]
\begin{center}
\begin{tabular}{|c||l|l|l|}
\hline
\rule{0pt}{15pt}
Law & \quad \, $d$~$=$~$3$ & \quad \, $d$~$=$~$4$ & \quad ~ $d$~$=$~$5$ \\ 
\hline
\rule{0pt}{15pt}
$\delta$ & $-~ 0.02(3)$ & $-~ 0.08(1)$ & $-~ 0.054(3)$ \\
$d_f$ & $-~ 0.02(4)$ & $-~ 0.082(16)$ & $-~ 0.054(2)$ \\
$\tau$ & $-~ 0.025(17)$ & $-~ 0.084(10)$ & $-~ 0.0547(25)$ \\
\hline
\end{tabular}
\caption{Estimates for $\eta$ derived from the scaling law of the indicated
exponent.}
\label{percetascal}
\end{center}
\end{table}}

The estimates for $\beta$, $d_f$ and $\gamma$, compiled from $13$ or $14$ 
acceptable approximants, have similar general properties in that the weighted 
means have tighter uncertainties reflecting the trend of the individual loop 
means. In order to provide balance on the use of scaling laws to extract 
individual exponent estimates, for example, we note that in three dimensions 
$d_f \Omega$~$=$~$1.51(3)$. Within uncertainties this is consistent with the 
bound valid for all dimensions of $d_f \Omega$~$\leq$~$\frac{3}{2}$ given in 
\cite{41}. For four and five dimensions the bound is comfortably satisfied. Our
Pad\'{e} estimates for $\omega$ also satisfactorily obey the same bound and 
within the uncertainty in three dimensions. Finally for $\nu$ we followed the 
same strategy as for $\delta$ here and in \cite{13} and constructed two-sided 
Pad\'{e}s for $1/\nu$. Estimates for $\nu$ are provided in Table \ref{percres} 
with those derived from $1/\nu$ marked as $\nu^\dagger$. For both means the
uncertainty on $\nu$ from using the series for $1/\nu$ was smaller. Given that 
we have around several dozen approximants for each of $\delta$ and $\nu$ we 
have combined the data for both and extracted additional estimates for these 
exponents. In other words we evaluate the accepted approximants for $1/\delta$ 
and $1/\nu$ in each of the three dimensions, compute their reciprocals before 
feeding these into the procedures used to compute $\mu$ and 
$\mu_{\mbox{\footnotesize{wt}}}$. The results of this exercise are provided in 
Table \ref{perccombres}. For the case of $\delta$ the combination is similar to
that for $1/\delta$ in terms of the central value and uncertainty. A similar 
observation is applicable for $\nu$.

{\begin{table}[ht]
\begin{center}
\begin{tabular}{|c|c||l|l|l|}
\hline
\rule{0pt}{15pt}
Exp & Ref \& Year & \quad ~ $d$~$=$~$3$ & \quad ~ $d$~$=$~$4$ & \quad  $d$~$=$~$5$ \\ 
\hline
\rule{0pt}{15pt}
$\alpha$ & \cite{13} 2021 & $-~ 0.64(4)$ & $-~ 0.75(2)$ & $-~ 0.870(1)$ \\
& This work & $\mathbf{-~ 0.704(14)}$ & $\mathbf{-~ 0.775(12)}$ & $\mathbf{-~ 0.875(4)}$ \\
\hline
\rule{0pt}{15pt}
$\beta$ & \cite{42} 1976 & $0.39(2)$ & $0.52(3)$ & $0.66(5)$ \\
& \cite{43} 1976 & $0.41(1)$ & --------- & --------- \\
& \cite{44} 1990 & $0.405(25)$ & --------- & --------- \\
& \cite{45} 1990 & --------- & $0.639(20)$ & $0.835(5)$ \\
& \cite{46} 2014 & $0.4180(10)$ & --------- & --------- \\
& \cite{24} 2015 & $0.4273$ & $0.6590$ & $0.8457$ \\
& \cite{47} 2021 & $0.4053(5)$ & --------- & --------- \\
& \cite{13} 2021 & $0.429(4)$ & $0.658(1)$ & $0.8454(2)$ \\
& This work & $\mathbf{0.425(15)}$ & $\mathbf{0.657(8)}$ & $\mathbf{0.845(2)}$ \\
\hline
\rule{0pt}{15pt}
$\delta$ & \cite{48} 1980 & $5.3$ & $3.9$ & $3.0$ \\
& \cite{49} 1997 & --------- & $3.198(6)$ & --------- \\
& \cite{50} 1998 & $5.29(6)$ & --------- & --------- \\
& \cite{13} 2021 & $5.16(4)$ & $3.175(8)$ & $2.3952(12)$ \\
& This work & $\mathbf{5.14(12)}$ & $\mathbf{3.17(3)}$ & $\mathbf{2.395(3)}$ \\
\hline
\end{tabular}
\caption{Summary of estimates for $\alpha$, $\beta$ and $\delta$ in various 
dimensions.} 
\label{percsumalbede}
\end{center}
\end{table}}

As no $\eta$ approximant passed the sieving process we have explored an
alternative to estimate this exponent. Examining the scaling laws of 
(\ref{percscal}) it is apparent that $\delta$, $d_f$ and $\tau$ solely depend 
on $\eta$. Therefore we have used their respective estimates for 
$\mu_{\mbox{\footnotesize{wt}}}$ in Table \ref{percres} to extract a measure of
$\eta$. The results are provided in Table \ref{percetascal} where the scaling 
law that was used is indicated in the first column. The three dimensional 
values have large relative uncertainties with those associated from the $\tau$ 
scaling law ensuring that the estimate is fully negative. For the other two 
dimensions the uncertainty gives a better refinement with those in five 
dimensions appearing to be the most reliable. Part of the reason why the direct
$\eta$ approximants are unreliable is that while the boundary values from six 
to two dimensions go from $0$ to $\frac{5}{24}$ the leading term of $\eta$ is 
negative for $\epsilon$~$\neq$~$0$. So as one approaches the lower dimensions 
any approximant would have to have a sharp change of slope to reach a 
relatively large value in two dimensions. This is not something which a 
rational polynomial is guaranteed to accommodate. Of the results using the 
three scaling laws it would appear that those from $\tau$ are the most reliable
partly as the uncertainty of the three dimensional estimate ensures it is 
negative in three dimensions.

Having discussed our analysis in detail we now place our exponent estimates in 
context by examining them in relation to results from other methods. Therefore
we have constructed Tables \ref{percsumalbede}, \ref{percsumdfet}, 
\ref{percsumganu}, \ref{percsumsita} and \ref{percsumomoo}. For each exponent 
these contain estimates in the three dimensions of interest in chronological 
order so that the progress in refining central values and uncertainties over 
half a century can be appreciated. It is not the case that the number of direct
evaluations is at the same level for all the exponents nor indeed for each of 
the three different dimensions. So central values for one exponent may not have
reached the level of commensurate accuracy as other exponents. In addition for 
each set of exponents we conclude with a boldface entry which are our final 
estimates from this study. Of great benefit in compiling previous results given
in these tables was the excellent source of \cite{51} which contains a live 
record of percolation exponents. We note that in Table \ref{percsumalbede} the 
estimate for $\beta$ associated with \cite{46} was derived from the scaling law 
using estimates of $d_f$ and $1/\nu$ in Tables \ref{percsumdfet} and 
\ref{percsumganu}. The conformal bootstrap estimate for $\nu$ from \cite{52} in
Table \ref{percsumganu} was deduced from a scaling law for the coupling of the 
energy operator in the underlying conformal field theory in that article. Also 
we included the constrained Pad\'{e} value for $\omega$ from \cite{13} in Table
\ref{percsumomoo} in order to compare with the present results.

{\begin{table}[H]
\begin{center}
\begin{tabular}{|c|c||l|l|l|}
\hline
\rule{0pt}{15pt}
Exp & Ref \& Year & \quad ~ $d$~$=$~$3$ & \quad ~ $d$~$=$~$4$ & \quad ~ $d$~$=$~$5$ \\ 
\hline
\rule{0pt}{15pt}
$d_f$ & \cite{53} 1985 & --------- & $3.12(2)$ & $3.69(2)$ \\
& \cite{49} 1997 & --------- & $3.0472(14)$ & --------- \\
& \cite{50} 1998 & $2.523(4)$ & --------- & --------- \\
& \cite{54} 1998 & $2.530(4)$ & --------- & --------- \\
& \cite{55} 2000 & $2.5230(1)$ & --------- & --------- \\
& \cite{56} 2001 & --------- & $3.046(7)$ & --------- \\
& \cite{57} 2001 & --------- & $3.046(5)$ & --------- \\
& \cite{58} 2005 & $2.5226(1)$ & --------- & --------- \\
& \cite{46} 2014 & $2.52293(10)$ & --------- & --------- \\
& \cite{24} 2015 & --------- & $3.0479$ & $3.528$ \\
& \cite{59} 2018 & --------- & $3.0437(11)$ & $3.524(2)$ \\
& \cite{52} 2018 & --------- & $3.003$ & --------- \\
& \cite{60} 2021 & --------- & $3.0446(7)$ & $3.5260(14)$ \\
& This work & $\mathbf{2.51(2)}$ & $\mathbf{3.041(8)}$ & $\mathbf{3.527(1)}$ \\
\hline
\rule{0pt}{15pt}
$\eta$ & \cite{45} 1990 & $-~ 0.07(5)$ & $-~ 0.12(4)$ & $-~ 0.075(20)$ \\
& \cite{49} 1997 & --------- & $-~ 0.0944(28)$ & --------- \\
& \cite{50} 1998 & $-~ 0.046(8)$ & --------- & --------- \\
& \cite{54} 1998 & $-~ 0.059(9)$ & --------- & --------- \\
& \cite{57} 2001 & --------- & $-~ 0.0929(9)$ & --------- \\
& \cite{24} 2015 & $-~ 0.0470$ & $-~ 0.0954$ & $-~ 0.0565$ \\
& \cite{13} 2021 & $-~ 0.03(1)$ & $-~ 0.084(4)$ & $-~ 0.0547(10)$ \\
& This work & $\mathbf{-~ 0.025(17)}$ & $\mathbf{-~ 0.084(10)}$ & $\mathbf{-~ 0.0547(25)}$ \\
\hline
\end{tabular}
\caption{Summary of estimates for $d_f$ and $\eta$ in various dimensions.} 
\label{percsumdfet}
\end{center}
\end{table}}

Taking a general overview of all the tables several themes are apparent. First 
there is a general trend over time of more precise values for exponents which 
can be associated with the improvement in computing technology. The evidence 
for this is the more accurate central values and tighter uncertainties. This is
particularly the case across each dimension for the exponents where there has 
been more focus such as $d_f$, $\nu$ and $\tau$. Aside from some of the earlier
years for these three examples there appears to be good agreement within 
uncertainties to two and sometimes three decimal places. In particular for 
$\nu$ and $\tau$ several different techniques were used to arrive at the 
recorded values. For the most part the Monte Carlo or high temperature results 
dominate the tables with a few estimates from conformal bootstrap or the 
functional renormalization group approach in addition to the present 
perturbative analysis. In this respect the results from the six loop 
constrained Pad\'{e} approximants are in close accord with previous four and 
five loop estimates. Where there is a subtle discrepancy in say the three 
dimensional exponents from the Pad\'{e} this might be due to using the two 
dimensional conformal field theory values for the two-sided approach. Any fixed
dimension computation will not have the freedom to bridge between discrete 
fixed spacetime dimensions. For other exponents clearly only five and six loop 
results are available for $\alpha$ but the slight discrepancy of perturbative 
estimates in three dimensions from results in the last thirty or so years is 
apparent in $\beta$ and $\delta$ for instance. For the remaining non-correction
to scaling exponents, aside from $\eta$ the three, four and five dimensional 
estimates from this six loop exercise, and lower loop ones, are comfortably 
within uncertainties from other techniques which is a reassuring observation 
indicative that the perturbative approach is independently competitive as a 
technique.

{\begin{table}[hb]
\begin{center}
\begin{tabular}{|c|c||l|l|l|}
\hline
\rule{0pt}{15pt}
Exp & Ref \& Year & \quad $d$~$=$~$3$ & \quad $d$~$=$~$4$ & \quad $d$~$=$~$5$ \\
\hline
\rule{0pt}{15pt}
$\gamma$ & \cite{42} 1976 & $1.80(5)$ & $1.6(1)$ & $1.3(1)$ \\
& \cite{43} 1976 & $1.6$ & --------- & --------- \\
& \cite{61} 1978 & $1.66(7)$ & $1.48(8)$ & $1.18(7)$ \\
& \cite{45} 1990 & $1.805(20)$ & $1.435(15)$ & $1.185(5)$ \\
& \cite{24} 2015 & $1.8357$ & $1.4500$ & $1.1817$ \\
& \cite{47} 2021 & $1.819(3)$ & --------- & --------- \\
& \cite{13} 2021 & $1.78(3)$ & $1.430(6)$ & $1.1792(7)$ \\
& This work & $\mathbf{1.81(4)}$ & $\mathbf{1.437(18)}$ & $\mathbf{1.180(3)}$ \\
\hline
$\nu$ & \cite{62} 1976 & $0.80(5)$ & --------- & --------- \\
& \cite{43} 1976 & $0.8(1)$ & --------- & --------- \\
& \cite{53} 1985 & --------- & --------- & $0.51(5)$ \\
& \cite{45} 1990 & $0.872(7)$ & $0.6782(50)$ & $0.571(3)$ \\
& \cite{49} 1997 & --------- & $0.689(10)$ & --------- \\
& \cite{50} 1998 & $0.875(1)$ & --------- & --------- \\
& \cite{55} 2000 & $0.8765(18)$ & --------- & --------- \\
& \cite{63} 2005 & --------- & --------- & $0.569(5)$ \\
& \cite{64} 2013 & $0.8764(12)$ & --------- & --------- \\
& \cite{65} 2014 & $0.8751(11)$ & --------- & --------- \\
& \cite{46} 2014 & $0.8762(12)$ & --------- & --------- \\
& \cite{24} 2015 & $0.8960$ & $0.6920$ & $0.5746$ \\
& \cite{66} 2016 & $0.8774(13)$ & $0.6852(28)$ & $0.5723(18)$ \\
& \cite{52} 2018 & --------- & $0.693$ & --------- \\
& \cite{67} 2020 & --------- & $0.6845(6)$ & $0.5757(7)$ \\
& \cite{13} 2021 & $0.88(2)$ & $0.686(2)$ & $0.5739(1)$ \\
& \cite{60} 2021 & --------- & $0.6845(23)$ & $0.5737(33)$ \\
& \cite{68} 2022 & $0.8762(7)$ & $0.6842(16)$ & $0.5720(43)$ \\
& This work & $\mathbf{0.895(20)}$ & $\mathbf{0.691(8)}$ & $\mathbf{0.5745(10)}$ \\
\hline
\end{tabular}
\caption{Summary of estimates for $\gamma$ and $\nu$ in various dimensions.} 
\label{percsumganu}
\end{center}
\end{table}}

The situation with our estimates of $\eta$ needs separate comment from the
other exponents. It is evident from Table \ref{percsumdfet} that the four, five
and six loop estimates are in the same ballpark. As noted earlier we used
scaling laws to derive the present values from other exponents which was the
method used at lower orders. So there the agreement is no surprise. Instead
there appears to be no overlap with results from other techniques. For instance
the central values from \cite{49} and \cite{57} are just about reached
by the lower end of the uncertainty band of the result from this work. No
parallel comment can be applied in the five dimensional case as only one value
is available from other methods. The situation in three dimensions closely
resembles that of four dimensions. Whether that could be explained by the use
of the two dimensional boundary condition cannot be ascertained for what is 
always a difficult exponent to measure accurately given its proximity to zero.
As a minor observation it is worth noting that all the earlier three 
dimensional estimates for $\eta$ in Table \ref{percsumdfet} are negative within
uncertainties which in one sense justifies our use of the $\tau$ scaling law 
for our $\eta$ estimate.

{\begin{table}[H]
\begin{center}
\begin{tabular}{|c|c||l|l|l|}
\hline
\rule{0pt}{15pt}
Exp & Ref \& Year & \quad $d$~$=$~$3$ & \quad $d$~$=$~$4$ & \quad $d$~$=$~$5$ \\
\hline
\rule{0pt}{15pt}
$\sigma$ & \cite{69} 1976 & $0.42(6)$ & --------- & --------- \\
& \cite{49} 1997 & $0.4522(8)$ & --------- & --------- \\
& \cite{60} 1998 & $0.445(10)$ & --------- & --------- \\
& \cite{46} 2014 & $0.4524(6)$ & --------- & --------- \\
& \cite{24} 2015 & $0.4419$ & $0.4742$ & $0.4933$ \\
& \cite{13} 2021 & $0.452(7)$ & $0.4789(14)$ & $0.49396(13)$ \\
& This work & $\mathbf{0.446(8)}$ & $\mathbf{0.476(5)}$ & $\mathbf{0.4934(11)}$ \\
\hline
\rule{0pt}{15pt}
$\tau$ & \cite{48} 1980 & --------- & $2.26$ & $2.33$ \\
& \cite{49} 1997 & $2.18906(8)$ & $2.3127(6)$ & --------- \\
& \cite{50} 1998 & $2.189(2)$ & --------- & --------- \\
& \cite{54} 1998 & $2.186(2)$ & --------- & --------- \\
& \cite{56} 2001 & --------- & $2.313(3)$ & $2.412(4)$ \\
& \cite{57} 2001 & $2.190(2)$ & $2.313(2)$ & --------- \\
& \cite{70} 2006 & $2.189(1)$ & --------- & --------- \\
& \cite{46} 2014 & $2.18909(5)$ & --------- & --------- \\
& \cite{24} 2015 & $2.1888$ & $2.3124$ & $2.4171$ \\
& \cite{59} 2018 & $2.1892(1)$ & $2.3142(5)$ & $2.419(1)$ \\
& \cite{13} 2021 & $2.1938(12)$ & $2.3150(8)$ & $2.4175(2)$ \\
& \cite{71} 2023 & --------- & --------- & $2.4177(3)$ \\
& This work & $\mathbf{2.194(4)}$ & $\mathbf{2.315(2)}$ & $\mathbf{2.4175(5)}$ \\
\hline
\end{tabular}
\caption{Summary of estimates for $\sigma$ and $\tau$ in various dimensions.} 
\label{percsumsita}
\end{center}
\end{table}}

Finally for the two correction to scaling exponents, $\omega$ and $\Omega$,
there does not appear to be a settled picture for the former. This is the one
case where the five and six loop perturbative estimates are indeed out of line
with the four loop one in three dimensions. However the reason for the
discrepancy is relatively simple. While all three methods employed the 
two-sided Pad\'{e} approximants a different two dimensional boundary condition 
was used in the four loop analysis of \cite{24} which was $2$ rather than the 
value of $\frac{3}{2}$ here and in \cite{13}. The former value from \cite{74}
has been superseded by the latter from \cite{41}. Curiously with the former 
value $\omega$ estimates were more in keeping with the central values of
\cite{50,55} whereas the higher loop order ones are not out of line with 
\cite{45}. For four and five dimensions the two dimensional constraint does not
seem to play a significant role in that the estimates from all approaches have 
a solid overlap within uncertainties. With regard to $\Omega$ there are only a
few non-loop based results for four and five dimensions and the loop estimates 
are generally in sink with them. In three dimensions aside from the results of 
\cite{54,59} the six loop estimate seems to be in the same company.

{\begin{table}[H]
\begin{center}
\begin{tabular}{|c|c||l|l|l|}
\hline
\rule{0pt}{15pt}
Exp & Ref \& Year & \quad \!\!\! $d$~$=$~$3$ & ~ $d$~$=$~$4$ & \quad  $d$~$=$~$5$ \\ 
\hline
\rule{0pt}{15pt}
$\omega$ & \cite{45} 1990 & $1.26(23)$ & $0.94(15)$ & $0.96(26)$ \\
& \cite{49} 1997 & --------- & $1.13(10)$ & --------- \\
& \cite{50} 1998 & $1.61(5)$ & --------- & --------- \\
& \cite{55} 2000 & $1.62(13)$ & --------- & --------- \\
& \cite{72} 2010 & --------- & $1.0(2)$ & --------- \\
& \cite{24} 2015 & $1.6334$ & $1.2198$ & $0.7178$ \\
& \cite{13} 2021 & $1.35(5)$ & $1.10(6)$ & $0.69(3)$ \\
& This work & $\mathbf{1.44(10)}$ & $\mathbf{1.18(9)}$ & $\mathbf{0.72(3)}$ \\
\hline
\rule{0pt}{15pt}
$\Omega$ & \cite{45} 1990 & $0.50(9)$ & $0.31(5)$ & $0.27(7)$ \\
& \cite{49} 1997 & --------- & $0.37(4)$ & --------- \\
& \cite{50} 1998 & $0.64(2)$ & --------- & --------- \\
& \cite{54} 1998 & $0.73(8)$ & --------- & --------- \\
& \cite{55} 2000 & $0.64(5)$ & --------- & --------- \\
& \cite{73} 2000 & $0.65(2)$ & --------- & --------- \\
& \cite{57} 2001 & $0.60(8)$ & $0.5(1)$ & --------- \\
& \cite{24} 2015 & --------- & $0.4008$ & $0.2034$ \\
& \cite{59} 2018 & $0.77(3)$ & --------- & --------- \\
& \cite{13} 2021 & --------- & --------- & $0.210(2)$ \\
& \cite{71} 2023 & --------- & --------- & $0.27(2)$ \\
& This work & $\mathbf{0.602(7)}$ & $\mathbf{0.407(5)}$ & $\mathbf{0.2089(11)}$ \\
\hline
\end{tabular}
\caption{Summary of estimates for $\omega$ and $\Omega$ in various dimensions.} 
\label{percsumomoo}
\end{center}
\end{table}}

\sect{Discussion.}

It is worth offering some general comments and overview of our study. First, we
have derived estimates for the critical exponents for two physics problems that
are governed by a scalar field theory with a cubic self-interaction by using 
the recently derived six loop renormalization group functions of \cite{14,15}. 
For practical applications the relevant exponent values are those in three, 
four and five dimensions which necessitates a resummation of the $\epsilon$ 
expansion. To ensure reliability of the lower dimensional values we extended 
the earlier two-sided Pad\'{e} calculations of \cite{13,24}. For both Lee-Yang 
and percolation theory the new estimates were not significantly dissimilar from
the lower loop ones indicating a degree of convergence and in a few cases an 
improvement on previous uncertainties using the weighted Pad\'{e} approximants.
While this is reassuring what is perhaps worth noting is that the latest 
estimates in general are not out of line with those by other methods. Although 
this has to be qualified in that it is they are in accord with results from 
more recent years such as the conformal bootstrap programme or the functional
renormalization group technique. With the improvement of computer technology 
and modern analytic techniques it appears evident that higher order 
perturbation theory can remain competitive. For instance in the Lee-Yang case 
the new fuzzy sphere method for three dimensions has close overlap with the 
loop results. The latter would not have been the case without the two-sided 
Pad\'{e} approximants since the two dimensional boundary condition on the 
rational polynomials was essential in shaping the monotonic behaviour of the 
exponents in $2$~$\leq$~$d$~$\leq$~$6$. From the point of view of higher order 
computations the graphical function method has been applied to $\phi^4$ theory 
to seven loops which is the current state of the art. In principle the 
renormalization of $\phi^3$ theory could be extended to the same order with 
that method. However to execute such a computation would probably correspond to
a significant increase in the level of difficulty for a scalar cubic theory.

\vspace{1cm}
\noindent
{\bf Data Availability Statement.} The data that support the findings of this
article are openly available \cite{75}.

\vspace{1cm}
\noindent
{\bf Acknowledgements.} The work was carried out with the support of the STFC 
Consolidated Grant ST/X000699/1. The author is grateful to O. Schnetz for
discussions and R.M. Ziff for useful correspondence on percolation exponents.
For the purpose of open access, the author has applied a Creative Commons 
Attribution (CC-BY) licence to any Author Accepted Manuscript version arising.

\end{document}